\documentclass[pdftex,twocolappendix,numberedappendix,appendixfloats,apj,twocolumn]{aastex63}
\usepackage{color}
\usepackage{amsmath}

\newcommand{\MH}{H$_2$\,}

\begin{document}
\shorttitle{RHD models of AGNs}

\shortauthors{Williamson, H\"{o}nig, \& Venanzi}

\title{Radiation hydrodynamics models of Active Galactic Nuclei: Beyond the central parsec}

\author{David Williamson}
\affil{Department of Physics \& Astronomy, University of Southampton, Southampton, SO17 1BJ, UK}
\email{d.j.williamson@soton.ac.uk}

\author{Sebastian H\"{o}nig}
\affil{Department of Physics \& Astronomy, University of Southampton, Southampton, SO17 1BJ, UK}
\email{s.hoenig@soton.ac.uk}

\author{Marta Venanzi}
\affil{Department of Physics \& Astronomy, University of Southampton, Southampton, SO17 1BJ, UK}
\email{m.venanzi@soton.ac.uk}

\begin{abstract}
We produce radiation hydrodynamics models of an AGN `torus' plus outflow on $1-100$ pc scales. This large scale permits direct comparison with observations, provides justification for configurations used in radiation transfer models, and tests the sensitivity of results of smaller scale dynamical models. We find that anisotropic radiation from an AGN accretion disk can cause an outflow to evolve to become more polar, agreeing with the ubiquity of polar extended mid-IR emission, and the general geometry predicted by radiative transfer models. We also find the velocity maps can reproduce many features of observations, including apparent `counter-rotation'.
\end{abstract}


\section{Introduction}\label{section_intro}

The `torus' is a key component of the classic unification picture of an active galactic nucleus (AGN) \citep{1993ARA&A..31..473A,2015ARA&A..53..365N}. Here, the AGN is a structure mostly illuminated by a bright accretion disk around a supermassive black hole (SMBH). Additional contributions to luminosity and extinction can come from other components, such as a disk wind \citep{1994ApJ...434..446K,1995ApJ...451..498M,2000ApJ...545...63E,2000ApJ...543..686P,2002ApJ...579..725L,2004ApJ...616..688P,2013A&A...551L...6K,2013MNRAS.436.1390H,2013MNRAS.435.3122G,2014MNRAS.438.3024L,2014ApJ...789...19H,2015MNRAS.450.3331M,2017MNRAS.467.2571M,2017MNRAS.471.4788M,2017MNRAS.465.2873N,2019ApJ...887..135L,2019MNRAS.482.5316M} or a warped disk \citep{1977ApJ...214..550P,1995ApJ...438..841P,1997MNRAS.292..136P,2005MNRAS.359..545N,2009MNRAS.398..535U,2010MNRAS.402..537W,2013MNRAS.431.2655K,2020MNRAS.492.5540M}. The luminous central engine is then obscured by an axisymmetric dusty molecular `torus', classically a static structure whose thickness is supported by infrared radiation pressure \citep{1992ApJ...399L..23P,1994MNRAS.268..235G,2007ApJ...661...52K,2008ApJ...679.1018S}. However, more recent observations and simulations suggest a more complex and dynamic picture.

High resolution ALMA observations show the molecular components of AGNs have signs of inflows, outflows, rotation, a possibly lopsided or misaligned morphology, and non-circular motions or perhaps even counter-rotation \citep{2016ApJ...823L..12G,2018ApJ...859..144A,2018ApJ...867...48I,2019A&A...628A..65A,2019ApJ...884L..28I,2019A&A...632A..61G,2019A&A...629A...6T}. Infrared interferometry additionally reveals an extended polar structure, suggesting a biconical dusty wind emitted from the `torus' region \citep{2012ApJ...755..149H,2013ApJ...771...87H,2014A&A...563A..82T,2016A&A...591A..47L,2018ApJ...862...17L}. These observations have spatial resolutions often approaching $\sim1$ pc, and reveal structure on $\sim10$ pc scales.

High resolution radiation hydrodynamics (RHD) simulations have confirmed that the torus could be supported by infrared radiation pressure, at least under certain conditions. The dust-gas mix is very opaque at UV wavelengths, and the radiation pressure (either UV or re-radiated IR) also drives an outflow \citep[][Paper I]{2011ApJ...741...29D,2012ApJ...747....8D,2012ApJ...761...70D,2012ApJ...758...66W,2015ApJ...812...82W,2016ApJ...825...67C,2016ApJ...819..115D,2016MNRAS.460..980N,2017ApJ...843...58C,2018MNRAS.473.4197C,2019ApJ...876..137W}. X-ray heating and supernovae can contribute to this simulated outflow and help inflate the torus \citep{2002ApJ...566L..21W,2016ApJ...828L..19W,2020ApJ...889...84K}. These sets of simulations all use different approaches to modelling RHD (and other physics), and produce a variety of outflow rates, as we illustrate further in Section~\ref{comparevels}.

At coarse resolutions, the UV radiation pressure is smoothed out over a large mass element, and the wind is suppressed (Paper I). Hence, to properly resolve the optically thick wind generation region, these simulations are usually performed at sub-parsec scale resolution, with domain sizes of $\sim1$ pc. This domain size contains the sublimation radius, where the effects of radiation on dust are at their most extreme.

Hence there is a gap between resolved simulations of the inner wind-generating region of the `torus', and observations of the `torus'. Often, the entire simulated domain is smaller than the resolution limit of the observations! These two scales may represent very different regions of the AGN -- an `inner torus' that is largely dominated by the radiation and gravity of the central engine, and an `outer torus' or `inner circumnuclear disk', where contributions from galactic physics (e.g. accretion, gravity potential), and interstellar medium physics (i.e. multi-phase and supernova-regulated) are more significant.

Therefore, there is motivation to produce a model that extends high resolution simulations to a larger physical domain, so that the results can be directly compared with observations. This gives insight as to whether the diversity in small scale `circum-sublimation-radius' simulation physics leads to a diversity in large-scale predictions, or whether the parameter space is degenerate and the sub-parsec physics can not be well constrained by super-parsec scale observations. This can also provide a theoretical justification for the wind+disk geometries required by radiation transfer simulations \citep{2017ApJ...838L..20H,2017MNRAS.472.3854S,2019MNRAS.484.3334S,2019ApJ...884...10G,2019ApJ...884...11G,2020ApJ...890..152M}. This approach can furthermore lead to an interpretation of the ambiguous results of observations, such as the conflicting explanations for apparently counter-rotating gas \citep{2016ApJ...823L..12G,2019ApJ...884L..28I}. Our dynamical model is based on the emerging picture that the `torus' is a combination of a disk, wind, and a puffed-up wind-launching region \citep{2019ApJ...884..171H}, with dynamics dominated by radiation pressure effects.

The layout of this paper is as follows: in Section~\ref{section_method} we explain our numerical method and the choice of simulation parameters; in Section~\ref{section_results}, we present the resulting properties and evolution of our simulations, as well as the results of mock observations; in Section~\ref{section_discussion}, we put these results in context of recent literature; and in Section~\ref{section_conclusion} we summarize the key discoveries of this work.

\section{Method}\label{section_method}

Our methodology is an extension of the model used in Paper I. That model consisted of smoothed-particle hydrodynamics (SPH) simulations of a parsec-scale disk of gas subject to heating and radiation pressure from a central AGN. In this work, we produce a larger scale model. This requires additional sub-grid physics to represent behavior that is now unresolved.

As illustrated in Figures 2 \& 3 in Paper I, a very fine resolution is required to correctly capture the process of radiation pressure generating winds at the inner edge of a  dense disk. The key criterion is that the flux-weighted optical depth of a resolution element must be $\tau\ll1$. This is a density dependent criterion, with the required resolution becoming finer at higher densities (i.e. higher volumetric opacities). Therefore, the effects of radiation pressure on an existing \textit{low density} wind can still be correctly captured at lower resolutions. It is only the \textit{generation} of winds -- or, at minimum, the uplifting of low density material -- that can not be accurately modelled at coarser resolutions.

Consequently, rather than explicitly modelling the production of winds, we use a sub-grid model that introduces uplifted material in the inner region with some injected velocity, and then explicitly evolve the low-density wind. The principle here is that it is better to be deliberately artificial than incorrect. We do not resolve the mechanism for uplifting material in the central region, so it is preferable to introduce a model that artificially generates uplifted material rather than attempting to `self-consistently' but inaccurately model wind generation. We emphasize again that it is only the {\em uplift} of material that is unresolved and must be generated with a sub-grid method -- the elevated gas has a low density, and can be accelerated rapidly by explicitly resolved radiation pressure. We can therefore produce winds even if the injection speed is well below escape velocity or even near zero.

In this Section we summarize the existing model and simulation code, and then explain the new sub-grid features. A more thorough description of the existing model is given in Paper I.

\subsection{Summary of existing model}\label{section_existing_model}

The core dynamical feature of the model is a dusty gas disk, impacted by an AGN radiation field. The opacity of dusty gas can be over a hundred times the Thomson opacity used to define the Eddington luminosity. Hence, an AGN with a luminosity as low as $1\%$ of its Eddington luminosity is well above the `dusty' Eddington luminosity, and in our simulations the radiation pressure produces dramatic outflows \citep[see][]{1992ApJ...399L..23P}. The anisotropic emission from an AGN accretion disk may allow gas to accrete along the accretion disk plane, but we find that this slice is too thin to be easily resolved in 3D Lagrangian simulations. Instabilities such as the Radiative Rayleigh-Taylor instability may provide another mode of dusty accretion \citep{2011ApJ...730..116J,2014MNRAS.437.2856P}, but the Rayleigh-Taylor instability is notoriously difficult to capture in SPH \citep{2007MNRAS.380..963A} (although as we note below, this effect is reduced in our simulations). Therefore, we focus our attention on the production and evolution of these outflows, where the accurate kinematics of SPH are most useful, and ignore accretion processes.

We also ignore the effects of magnetic fields. Magnetic fields may have some role in angular momentum transport within the thin disk and in launching the wind \citet{1991ApJ...376..214B,1994ApJ...421..163B,2017ApJ...843...58C,2018A&A...615A.164V}, but these phenomena are largely subsumed into the `sub-grid' region of our simulations. One radiation magnetohydrodynamics study found that radiation dominates the outflow \citep{2017ApJ...843...58C}, while an analytic calculation in the strong-field limit (i.e. equipartition between thermal and magnetic energy densities) found that magnetic and radiative energy densities are similar in the disk, but thin outflowing gas is dominated by radiation pressure \citep{2018A&A...615A.164V}. It is also unclear to what extent idealized MHD is appropriate for modelling polarised dusty molecular gas.

Hydrodynamics and gravity are solved with GIZMO \citep{2015MNRAS.450...53H} in P-SPH mode, using an ideal gas equation of state. The P-SPH formalism was used in the first set of FIRE simulations \citep{2014MNRAS.445..581H}, and is a `modern' SPH formalism that resolves or reduces many of the traditional problems with SPH \citep{2013MNRAS.428.2840H}. This includes improved gradient calculations and improved artificial viscosity allowing better capturing of interfaces and instabilities, and reduction of spurious shear, although P-SPH still smooths out shocks when compared with grid and Godunov-style schemes \citep{2015MNRAS.450...53H}. As much of the gas in our simulations is cold with low sound speeds ($\sim1$\,km\,s$^{-1}$) compared with typical flow speeds ($\sim100$s of km\,s$^{-1}$), we do expect shocks to be present in colliding flows. However, while these problems may appear dramatic in idealized tests, comparisons of galaxy simulation codes have found the differences are dominated by sub-grid modelling rather than hydrodynamic schemes \citep{2012MNRAS.423.1726S,2015MNRAS.450...53H,2015MNRAS.454.2277S}. Similarly, here we are dominated by radiation effects.

In our code, radiative accelerations, flux-weighted opacities, and radiative heating and cooling rates, and are calculated with a series of \textsc{Cloudy} models \citep{2013RMxAA..49..137F,2017RMxAA..53..385F}, and tabulated for interpolation in the simulation. We also tabulate the emissivities of a selection of lines for post-processing analysis, but these do not affect the dynamics of the simulation.

The \textsc{Cloudy} models include an AGN SED, a constant stellar background of $1000$ Habing units, a dust grain model that includes graphites and silicates with sizes following the MRN distribution \citep{1977ApJ...217..425M}, and extinction from dust and gas (including molecules). The central AGN engine is assumed to be the only (non-background) source of radiation. We varied the Eddington factor in Paper I, but in this work we almost always use a fixed value of $\gamma_{edd}=0.01$.

We produce three tables: a `dusty' table, a `sputtered' table, and a `high density' table. The `dusty' and `high density' tables include grain molecules, while the `sputtered' table does not. The sputtered table is used for particles with temperatures $10^5-10^8$ K, the `high density' table is used for particles with temperatures $10^1-10^3$ K and densities $10^8-10^9$\,cm$^{-3}$, and the `dusty' table is used otherwise. We interpolate between the tables where appropriate. The high-density table only contains cool-warm gas. Hot dense gas does not exist in our simulations, and it was difficult for the \textsc{Cloudy} runs to converge at these high densities. Hence we were justified in not performing superfluous \textsc{Cloudy} runs for an empty part of the phase diagram.

The main `dusty' table includes $8$ densities from $10^0-10^7$\,cm$^{-3}$, equally spaced in log space. The temperature from $10^1-10^5$ is tabulated in $17$ equally log-spaced steps. Radiation intensity is tabulated from $10^{1.25}-10^{9.25}$ erg\,cm$^{-2}$\,$s^{-1}$ in $17$ equally log-spaced steps. Each \textsc{Cloudy} run outputs a large number of cells at different optical depths, which we interpolate to $50$ equally log-spaced steps from $\tau=0.01$ to $\tau=7$.

We note that some plots in this paper appear to have gas with $n_H$ below $10^0$\,cm$^{-3}$. In the zoomed-in plots ($\leq20$ pc), this is an artifact of the smoothing kernel used in the plots, and not representative of the particle densities used in the simulation. There is `real' low density gas in the simulations, but only at large distances where the particles are not dynamically important. Examining one sample snapshot (heavy\_vesc\_55\_thick at $\sim2$ Myr), we find that $99.5\%$ of gas with $n_H<10^0$\,cm$^{-3}$ is beyond $20$ pc, and $77\%$ is beyond $100$ pc. Similarly, the snapshot has no gas hot enough to use the `sputtered' table. Some particles ($\sim2\%$ in the sample) do exceed the tabulated radiation intensity, in which case we extrapolate by assuming that heating and radiative acceleration increase proportionally with intensity. 

We continue to use an SMBH of mass $M_{BH}=10^6$ M$_\odot$. These are on the low end of AGN masses and Eddington factors, and are chosen as a compromise with computational expense. Higher masses and luminosities increase accelerations and require a shorter time step. A higher mass SMBH would require a higher mass `torus', and a larger particle count -- increasing the particle mass too much has an effect on radiation pressure, as noted above. Hence we model an AGN with a luminosity of $L\sim10^{42}$~ergs$^{-1}$, weaker than typically observed AGNs \citep{2002ApJ...579..530W,2010A&A...516A..87S}. We do perform one run with $\gamma_{edd}=0.08$ to test the luminosity dependence (section~\ref{section_highedd}). We do not self-consistently calculate accretion of mass and angular momentum onto the SMBH accretion disk, and so we keep the AGN luminosity constant within each simulation. 

In the dynamical simulations, AGN radiation is emitted anisotropically, following

\begin{equation}
F(r,\phi)=\frac{L}{4\pi r^2} f(\phi)
\end{equation}
where $L$ is the luminosity of the AGN, $r$ is the distance from the AGN, $\phi$ is the angle from the equatorial axis of the AGN, and the anisotropy function $f(\phi)$ is defined to be
\begin{equation}
f(\phi)=\frac{1+a\sin\phi+2a\sin^2\phi}{1+2a/3}
\end{equation}
where we define $a=(\eta_a-1)/3$, and define $\eta_a$ as the ``anisotropy factor'', equal to the ratio between the polar flux and the equatorial flux. This is a free parameter to control the relative strengths of disk emission and isotropic emission, and can be varied between the simulations to represent different models of the central engine.

Optical depths from the AGN to each particle are calculated with a raytracing method, using an oct-tree algorithm \citep{Revelles00anefficient} to efficiently detect collisions between rays and SPH particles.

The simulation includes a fixed background potential consisting of a Plummer-softened Keplerian component representing the SMBH, and a Hernquist component to represent a central stellar bulge (of mass $M_H=10^9$ M$_\odot$, and scale $c_H=250$ pc). This choice of bulge mass comes from the $M-\sigma$ relation, which implies a roughly 1000:1 ratio between SMBH mass and bulge mass \citep{2004ApJ...604L..89H,2013ApJ...764..184M}. In the simulations in this work, the Plummer softening length is $h_{BH}=0.01$ pc, larger than that of Paper I.

Self-gravity is also included. We initalize the disk with a rotation curve that is balanced with both the imposed potential and the gas.

Combined with radiative cooling, self-gravity can cause the gas to collapse into dense, potentially star-forming, clumps. Gas particles above a threshold density are considered `star-forming'. Each time-step, these particles may be deleted from the system with a probability corresponding to the local free-fall time-scale. In the simulations in this work, the mass of star formation is much less than the bulk mass of the gas, and far less than the mass of the bulge and SMBH. Hence we do not track the newly formed stars, but instead assume they form a negligible portion of the mass of the stellar bulge.

\subsection{Additions and modifications to the model}\label{section_modifications}

\begin{figure}
\begin{center}
\includegraphics[width=\columnwidth]{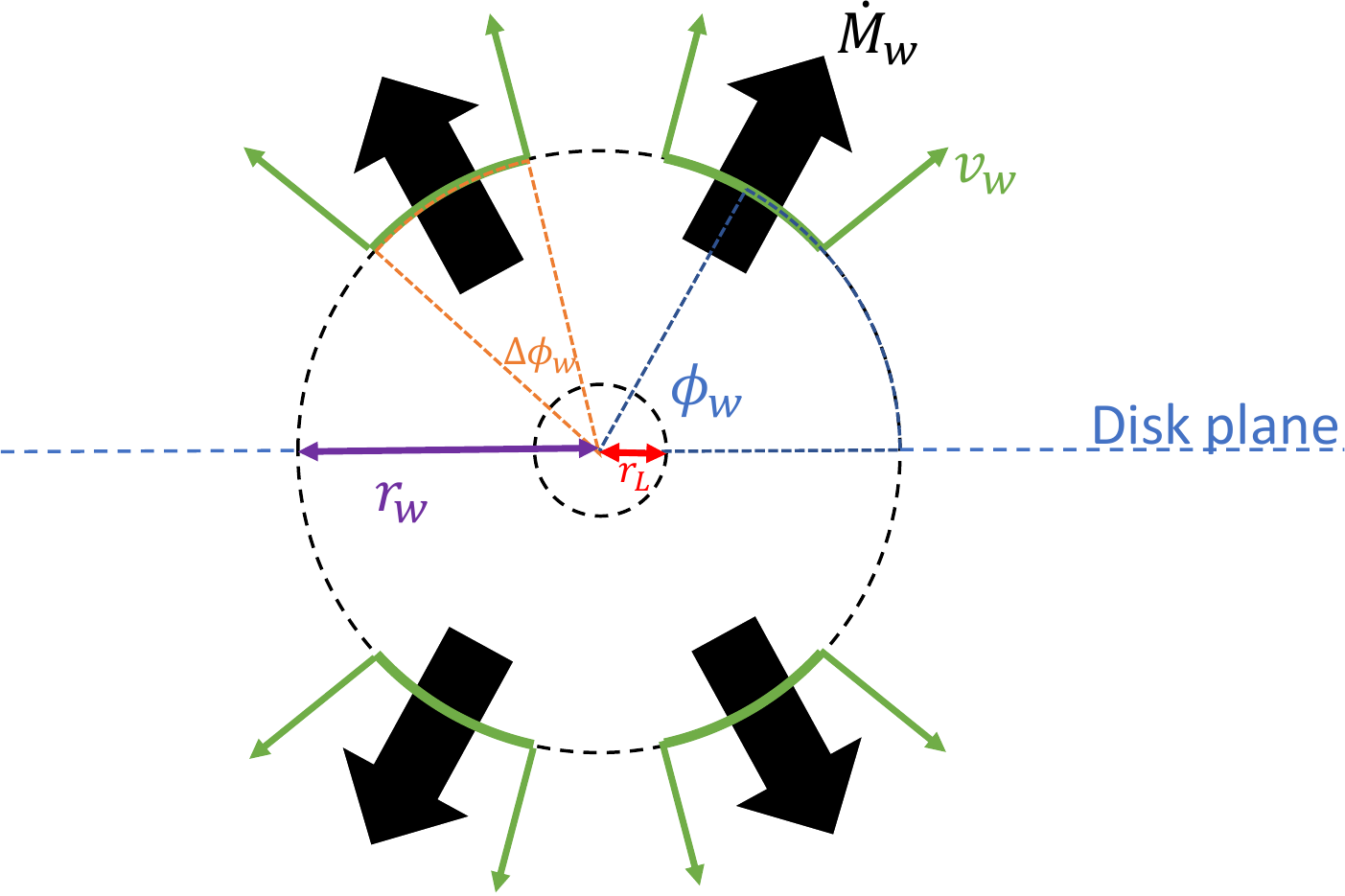}
\end{center}
\caption{\label{windgen}
A schematic of the injected wind parameters, as described in Section~\ref{section_modifications}. Note that $\dot{M}_w$ is the mass outflow rate for the entire wind -- not for one sector.}
\end{figure}

Particles are produced across a range of inclinations on the surface of a sphere centered on the origin, axisymmetrically and evenly distributed. This produces a biconical outflow. The generation of this material is governed by a number of parameters, which we list here, where the `w' sub-script stands for ``wind': the mass generation rate ($\dot{M}_w$), the initial radial velocity ($v_w$), the radius where winds are injected ($r_w$), the inclination of the center of the generation region ($\phi_w$), the (unresolved) wind launching radius ($r_L$) for determining the initial tangential velocities, and the thickness of the generation region ($\Delta\phi_w$). These properties are illustrated in Figure~\ref{windgen}. Tangential velocities are set by assuming that the wind particles conserve specific angular momentum from Keplerian circular motion at $r_L$ to particle generation at $r_w$. That is, the tangential velocity of the wind particles is $v_\perp = \sqrt{GMr_L}/r_w$.

We also include a sub-grid model for unresolved extinction in the central region. This is the unresolved ``inner torus'' region where internal radiative transfer effects (such as vertical support from infrared radiation pressure) are more significant. Sub-grid extinction is assumed to be axisymmetric. The optical depth follows a split linear function -- equivalently, we could say (with fixed mass opacity) the column density follows a split power law. This is given by:

\begin{equation}
\tau=
\begin{cases} 
      \tau_0 - \phi/\alpha_E & \phi\leq \phi_T \\
      \tau_0 - \phi_T/\alpha_E - (\phi-\phi_T)/\beta_E & \phi\geq \phi_T 
   \end{cases}
\end{equation}

where $\tau_0$ is the optical depth along the inner torus plane, $\alpha_E$ is the optical depth slope for the dense equatorial part of the inner torus, $\beta_E$ is the optical depth slope for the higher inclination region of the inner torus, and $\phi_T$ is the inclination of the transition between these two regimes. This type of profile is produced in our small-scale simulations in Paper I. Similar profiles are found in other dynamical simulations \citep[e.g.][]{2015ApJ...812...82W}. We use a larger $\phi_T$ and a lower $\alpha_E$ than predicted by the models in Paper I, to represent the thicker `torus' we expect from infrared radiation pressure, and from the high covering fractions found in observations \citep[e.g.][]{2013ApJ...777...86L,2013MNRAS.429.1494R}.

\subsection{Simulations}\label{section_simulations}

We produce a suite of runs, varying the anisotropy of the input radiation field, the sub-grid extinction profile, the wind mass input rate, the initial velocity of the input wind, and the range of inclinations of wind particles. In all runs, we inject the wind at $r_w=1$ pc. Wind and disk particles have a mass of $0.1$\,M$_\odot$. The disk is initalized with $10^6$ particles, giving $M_d=10^5$\,M$_\odot$. Observations have shown that a `torus' typically has a mass on the order the SMBH's \citep{2016ApJ...823L..12G,2018ApJ...859..144A,2018ApJ...867...48I}, and we select a disk mass on the small end of this range, for computational efficiency.

At the fiducial luminosity, the AGN has a sublimation radius of $r_{sub}=0.015$. Based on previous simulations, we assume that the critical wind launching region is around the sublimation radius \citep[][Paper I]{2016ApJ...825...67C,2016ApJ...819..115D,2016MNRAS.460..980N,2019ApJ...876..137W}. Hence we set the sub-grid wind-launching radius to $r_L=r_{sub}=0.015$ pc. The tangential velocities of the wind set by conservation of angular momentum from $r_L$ (see Section~\ref{section_modifications}). We note that $r_w\approx70r_{sub}$, and that most of the details of wind generation should be contained in the sub-grid region $r<r_w$. 

We select three characteristic initial wind velocities to investigate: $10$ km s$^{-1}$, $100$ km s$^{-1}$, and $500$ km s$^{-1}$. We can characterize these velocities by comparison with the escape velocity. The escape velocity at $r=1$ pc from our gravitational potential is $\sim207$ kms$^{-1}$, and so these choices represent $v\ll v_{esc}$, $v\sim v_{esc}$ and $v \gg v_{esc}$. The slow $v=10$ km s$^{-1}$ winds represent gas that has sloshed out of the inner regions of the system but is not yet a strong wind. These runs test the acceleration of gas in the $r\ge1$ pc region. The $v=100$ km s$^{-1}$ runs are in the the regime where the wind may or may not fail, depending on the contributions of radiation pressure and self-gravity from the gas. For the $v=500$ km s$^{-1}$ runs, $v\gg v_{esc}$ and the dynamical situation becomes simple -- the wind trajectory is mostly ballistic. Hence we mostly focus our analysis on lower velocities.

We select two sets of outflow rates: $0.0546$ M$_\odot$yr$^{-1}$, and $0.546$ M$_\odot$yr$^{-1}$ for the wind-test runs (section~\ref{section_shortruns}), and $0.0378$ M$_\odot$yr$^{-1}$, and $0.378$ M$_\odot$yr$^{-1}$ for the full-scale runs (section~\ref{section_fullruns}). The wind covers a smaller solid angle in the full-scale runs, and the lower mass input rate means that the input mass outflow rate per steradian is closer between the two suites of runs. In particular, the large-scale runs with $\phi_w=35^\circ$ and $\Delta\phi_w$ have the same input mass outflow rate per steradian as the small-scale runs. The precise value of $0.546$ (as opposed to a round number like $0.5$) is somewhat arbitrary -- it is produced from a round number in test runs not published here.

The luminosity of our fiducial simulations corresponds to a SMBH accretion rate of
\begin{equation}
\dot{M}_{in} = \frac{L}{c^2 \eta} \approx 2.2\times10^{-4}\,\mathrm{M}_\odot \mathrm{yr}^{-1} \left(\frac{\eta}{0.1}\right)^{-1},
\end{equation}
where $\eta$ is the radiative efficiency of accretion. Hence, we are assuming that the majority of inflowing gas is ejected as a wind, and only a fraction is accreted.

Given the different mass outflow rates and injected wind speeds, the mechanical luminosities of our injected winds ($\dot{m}v_w^{2}/2$) range from $10^{-6}L$ to $0.034L$, representing a large range of coupling factors.

The outflow velocities selected for the models in this paper span the lower range of outflow velocities found in smaller scale RHD simulations (\citealt{2016ApJ...819..115D,2016ApJ...825...67C,2016MNRAS.460..980N,2017ApJ...843...58C}; paper I, also see Section~\ref{comparevels}), as this is where dynamical simulations of the outflow are most insightful. If $v_{w}\gg v_{esc}$, then the outflow will simply produce a cone, provided that ISM structure is ignored as we do in this paper. This simple geometry could be generated analytically for a radiation transfer code.

We use four models for the radiation field and sub-grid opacities, summarized in Table~\ref{radsum}. `defaultaniso' and `thinaniso', represent an anisotropic radiation source with a sub-grid anisotropic obscuring `torus' of two different thicknesses. To test to what extent the outflow properties are affected by the anisotropy of the radiation source and obscuration, we also test a model with isotropic obscuration (`iso'), and a model with isotropic obscuration and emission (`doubleiso'). 

\begin{table}
\begin{center}
\begin{tabular}{rccccc}
\hline\hline
Name & $\tau_0$ & $\alpha_E$ & $\phi_T$ & $\beta_E$ & $\eta_a$ \\
\hline
defaultaniso & $5$ & $10$ & $40$ & $50.5$ & $0.01$\\
thinaniso & $1$ & $50$ & $40$ & $252.5$ & $0.01$\\
iso & $1$ & $\infty$ & n/a & $\infty$ & $0.01$\\
doubleiso & $1$ & $\infty$ & n/a & $\infty$ & $1$\\
\end{tabular}
\end{center}
\caption{\label{radsum} \textup{Summary of input radiation parameters}}
\end{table}

\section{Results}\label{section_results}

\subsection{Wind-test runs}\label{section_shortruns}

\subsubsection{Wind-test runs -- setup of simulations}\label{section_shortruns_setup}

We perform a suite of $24$ wind-test runs, to investigate the self-consistency of outflows. These models have a run-time of $100$~kyr. The orbital time at $1$ pc (the wind injection radius) is $t_{orb}\approx15$~kyr, increasing to $t_{orb}\approx140$~kyr at a radius of $5$ pc. Hence, these short simulations can test the evolution of the wind near the injection radius, but can not show the evolved configuration on (e.g.) a $10$ pc scale.

In the wind-test runs, the outflows are generated isotropically (i.e. across the entire surface of a sphere, $\phi_w=\Delta\phi_w=45^\circ$) to test which inclinations produce a successful wind. However, the AGN emits radiation anisotropically (section~\ref{section_existing_model}), and we explore all four sub-grid opacity models (table~\ref{radsum}), some of which are also anisotropic. This means the resulting wind can evolve to become anisotropic. Along with two different mass outflow rates and three different injection speeds $v_w$, this comes to a total of $2\times3\times4=24$ wind-test runs.

\subsubsection{Wind-test runs -- results}\label{section_shortruns_results}

\begin{figure}
\begin{center}
\includegraphics[width=\columnwidth]{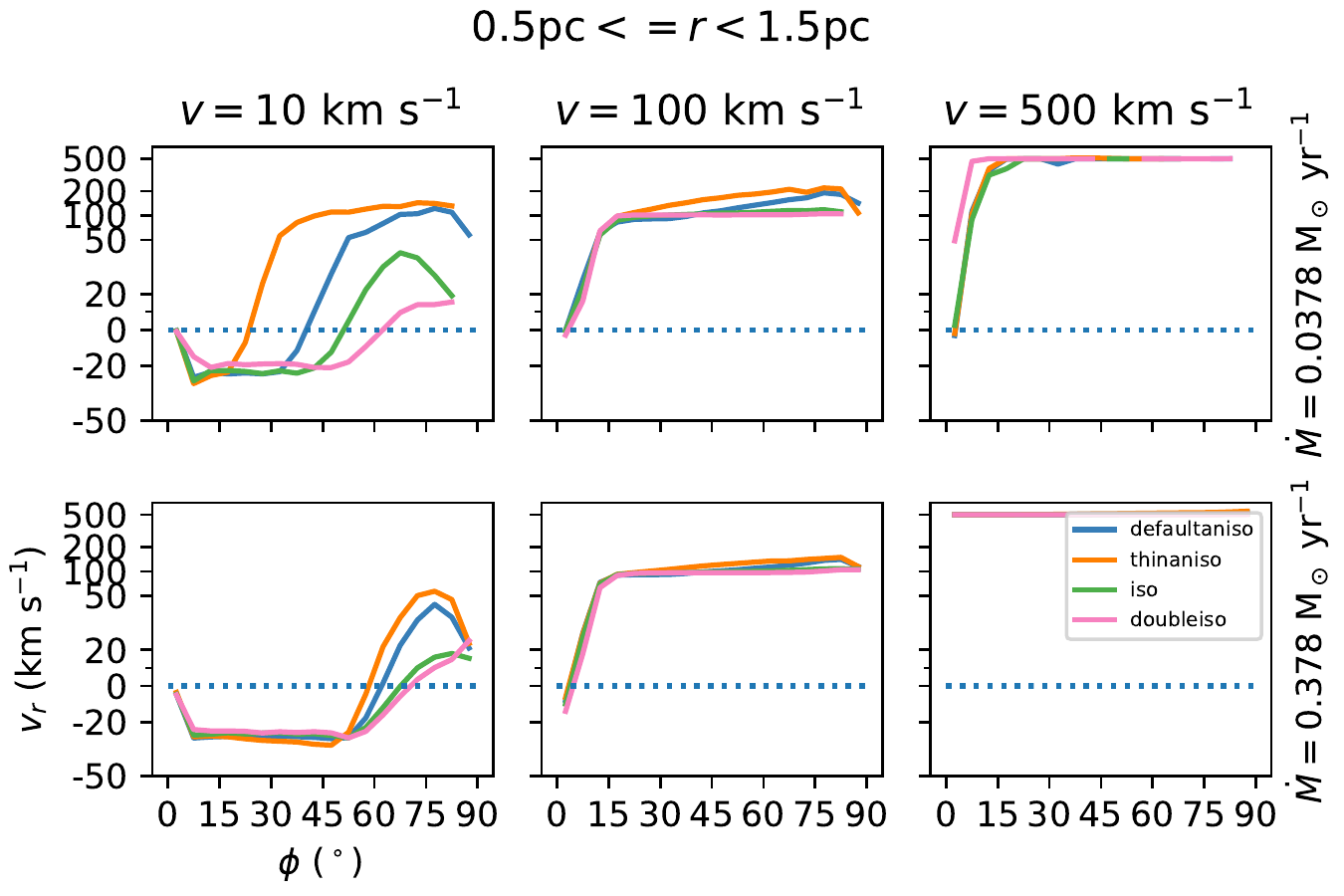}
\includegraphics[width=\columnwidth]{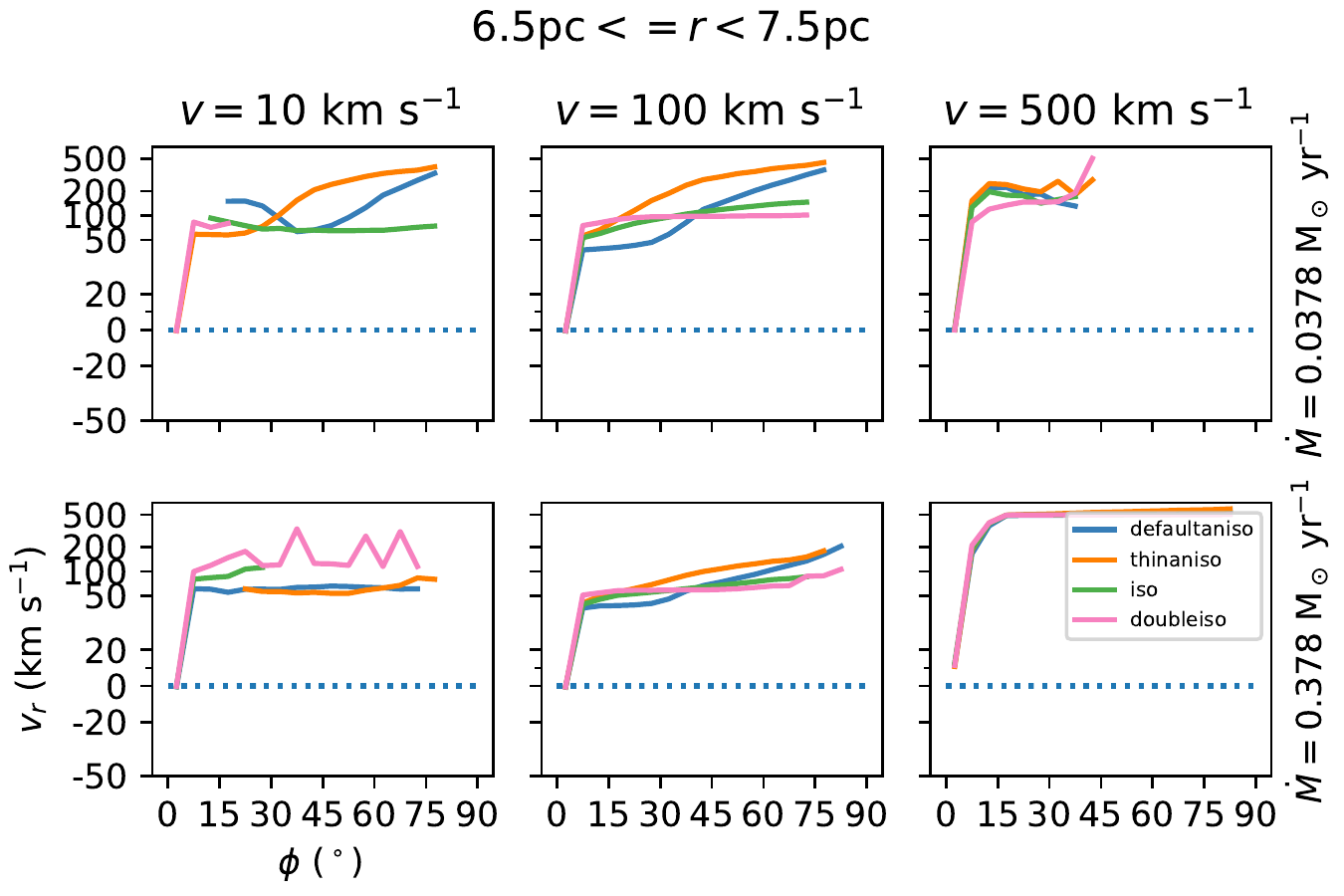}
\end{center}
\caption{\label{test_outflow_v}
Mean outflow velocities as a function of inclination for all test runs at $t=100$ kyr, in spherical shells of radius $r=1$ pc and $r=7$ pc.}
\end{figure}

We take snapshots from the end of the test simulations, and measure the angle-dependent mass-weighted radial velocities in spherical shells to characterize the outflow. We plot the results for two spherical shells in Figure~\ref{test_outflow_v} -- an inner shell at $0.5$ pc $\le r<1.5$ pc, and an outer shell at $6.5$ pc $\le r<7.5$ pc. The inner shell contains the wind generation region, but only a somewhat successful wind can reach the outer shell. A negative radial velocity ($v_r<0$) in the inner shell indicates the gas is unlikely to have reached the injected speed, radius, and inclination -- i.e. we should not inject wind of this speed at this inclination in the coming large-scale simulations. A negative radial velocity in the outer shell indicates a failed wind or fountain -- i.e. the gas has flowed out to reach this radius or beyond, but is now falling back inwards. The exact configuration of this outer bin is not converged with time, but is qualitatively indicative of the results expected in the full scale runs.

We find $v_r<0$ at low inclinations ($\phi<10^\circ$) in the inner region in almost all runs -- this is simply the wind hitting the disk. The only exception are the $\dot{M}=0.378$ M$_\odot$ yr$^{-1}$, $v_w=500$ km s$^{-1}$ runs, where the large outwards momentum destroys the disk. As noted above (section~\ref{section_simulations}), these high-speed winds otherwise flow radially outwards, largely unaffected by RHD forces, and barely even by gravity over these short time-scales.

For the other runs, the outflow is accelerated significantly by radiation pressure, and winds are ubiquitous, even when the injection speed $v_w\ll v_{esc}$. The acceleration is stronger when the outflow rate is lower, and when the sub-grid extinction is lower, but also increases as the radiation field becomes more anisotropic. This is analogous to super-Eddington accretion. Focusing flux in a smaller solid angle can provide enough radiation pressure to overcome gravity and accelerate the outflow, even if the isotropically averaged flux could not. For the $v_w=100$ km s$^{-1}$ runs, the outflows can continue to accelerate out to the outer shell, reaching a maximum of $v_r\sim500$ km s$^{-1}$.

Even though the radiation pressure is entirely radial, its strength is weaker at lower inclinations (closer to the disk plane). This means that winds at lower inclinations are more likely to fail, while winds at higher inclinations tend to be accelerated. The result is that the outflow tends to flow ``upwards'' more rapidly, and to flow in equatorial directions more slowly, or to even fail.

Hence the initially uniform outflow velocity also becomes more polar. This effect is stronger at lower mass outflow rates, which can be accelerated more efficiently by radiation pressure. The speed and covering angle of the vertical outflow also increases with increasing anisotropy of the radiation field: the outflow of `doubleiso' is weaker than that of `iso', which is weaker than `thinaniso' and `doubleaniso'. These last two only differ in the scaling of their sub-grid extinction, and so `thinaniso' produces a stronger outflow than `doubleaniso', simply because (after sub-grid extinction), it is a stronger radiation field everywhere.

Overall, we have found that outflows are successful in all runs for at least some range of angles, even if the wind has a near-zero injection speed. But the most critical point is that the generation of polar winds appears to be robust. The anisotropy of radiation pressure from the AGN accretion disk can allow more equatorial flows to fail. Hence, even if the wind at small scales is isotropic, larger-scale evolution can cause it to become more vertical. This means that the geometry of the large-scale structure is not strongly dependent on the initial properties of wind produced near the sublimation radius.

\subsection{Full-scale runs}\label{section_fullruns}

\begin{table}
\begin{center}
\begin{tabular}{rcccc}
\hline\hline
Name & $\dot{M}$ & $v_w$ & $\phi_w$ & $\Delta \phi_w$ \\
~ &   (M$_\odot$ yr$^{-1}$) & km s$^{-1}$ ($^\circ$) & ($^\circ$)\\
\hline
heavy\_slow\_55\_thick & $0.3780$ & $ 10$ & $55^\circ$ & $50^\circ$\\
heavy\_slow\_35\_thick & $0.3780$ & $ 10$ & $35^\circ$ & $50^\circ$\\
heavy\_vesc\_55\_thick & $0.3780$ & $100$ & $55^\circ$ & $50^\circ$\\
heavy\_vesc\_35\_thick & $0.3780$ & $100$ & $35^\circ$ & $50^\circ$\\
light\_slow\_70\_thin & $0.0378$ & $ 10$ & $70^\circ$ & $20^\circ$\\
light\_slow\_55\_thick & $0.0378$ & $ 10$ & $55^\circ$ & $50^\circ$\\
light\_slow\_35\_thick & $0.0378$ & $ 10$ & $35^\circ$ & $50^\circ$\\
light\_vesc\_70\_thin & $0.0378$ & $100$ & $70^\circ$ & $20^\circ$\\
light\_vesc\_55\_thick & $0.0378$ & $100$ & $55^\circ$ & $50^\circ$\\
light\_vesc\_45\_thin & $0.0378$ & $100$ & $45^\circ$ & $20^\circ$\\
light\_vesc\_35\_thick & $0.0378$ & $100$ & $35^\circ$ & $50^\circ$\\
light\_vesc\_20\_thin & $0.0378$ & $100$ & $20^\circ$ & $20^\circ$\\
light\_rapid\_55\_thick & $0.0378$ & $500$ & $55^\circ$ & $50^\circ$\\
light\_rapid\_35\_thick & $0.0378$ & $500$ & $35^\circ$ & $50^\circ$\\
\hline
\end{tabular}
\end{center}
\caption{\label{biginit} \textup{Summary of initial conditions of full-scale runs}}
\end{table}

Based on the results of these wind-test runs, we produce a series of longer ($t>1$ Myr) runs, that can be examined in a more steady state. These are no longer isotropic winds, but are injected at across some range of inclinations in both hemispheres, to represent biconical outflows produced at the inner face of the `torus'. We vary the input speed and outflow mass rate as in the small scale simulations, but also vary the thickness and angle of the wind. In particular, we include both equatorial and polar wind launching angles. This greatly increases the parameter space, and so we restrict the sub-grid opacity model to the doubleaniso model. The wind parameters for these simulations are listed in Table~\ref{biginit}. We perform $14$ simulations in total.

The run-times range from $1-3$ Myr, depending on when the morphology has reached a steady state. The `rapid' runs reach a steady state more quickly, as the gas flow is largely radial, and only simple structure is formed. Some of the `slow' runs also reach a steady state quickly, as the gas that does escape is accelerated vertically to $v\gg v_{esc}$. For these runs, we terminate the simulation after $t\sim1$ Myr. But in most of the `vesc' simulations, and in some of the `settled' simulations with more equatorial flows, a lot of gas can fall back in, creating a more complex structure. As a rough analytic estimate of the time-scale of this, a radial wind launched at $100$~km\,s$^{-1}$ from $r=1$~pc in our potential would reach a distance of $r_{max}=16$~pc, where the dynamical time ($\sqrt{r_{max}/|a_{grav}(r_{max})|}$) is $t_{dyn}\sim440$~kyr. We must complete several of these `cycles' to reach a steady state, requiring simulation times of $2$--$3$~Myr.

\subsubsection{Full-scale runs -- development of extended outflow, fountain \& `torus'}\label{fountain_results}

\begin{figure*}
\begin{center}
\includegraphics[width=.95\textwidth]{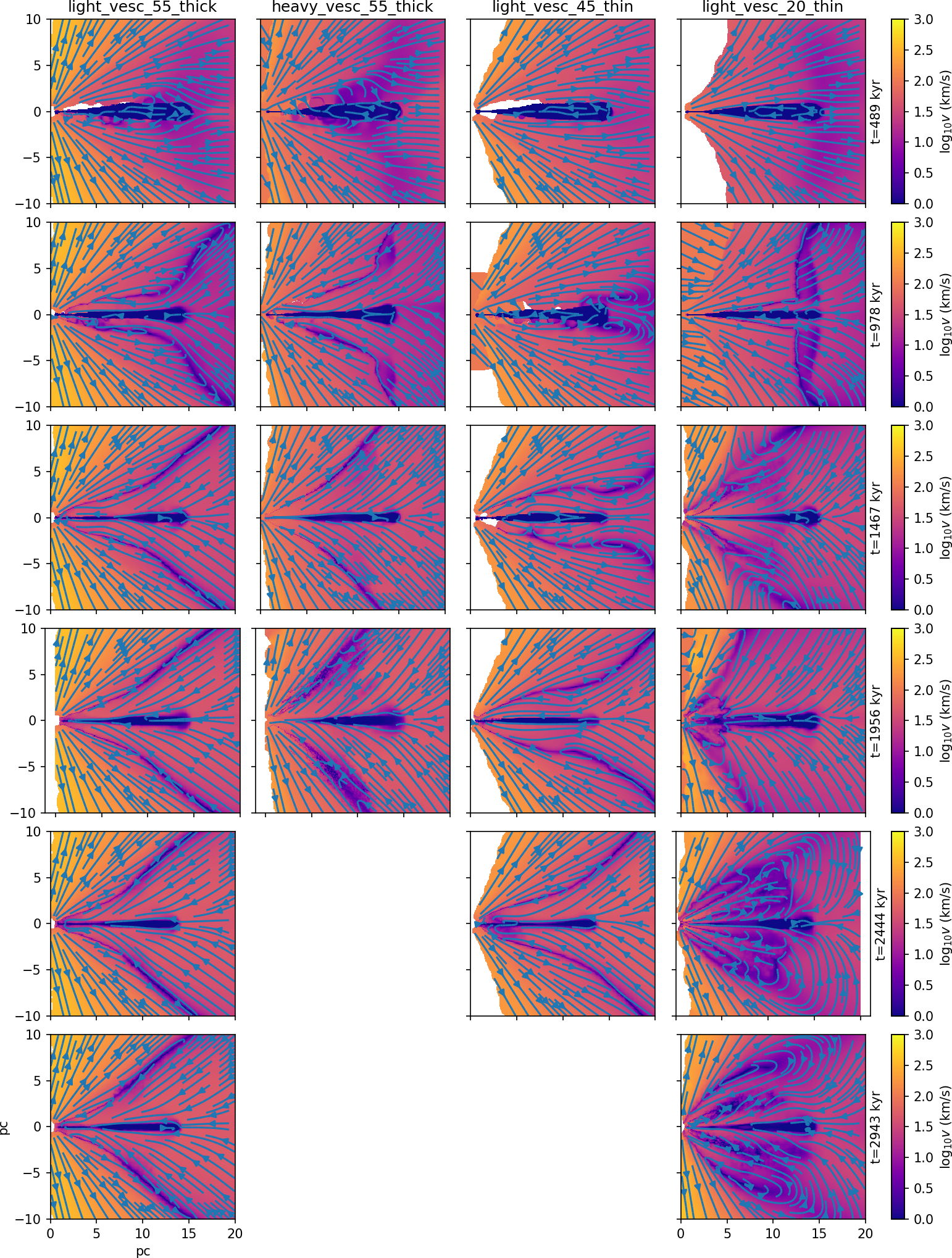}
\end{center}
\caption{\label{evol_v}Evolution of azimuthally averaged velocity fields of four sample full-scale runs }
\end{figure*}

\begin{figure*}
\begin{center}
\includegraphics[width=.95\textwidth]{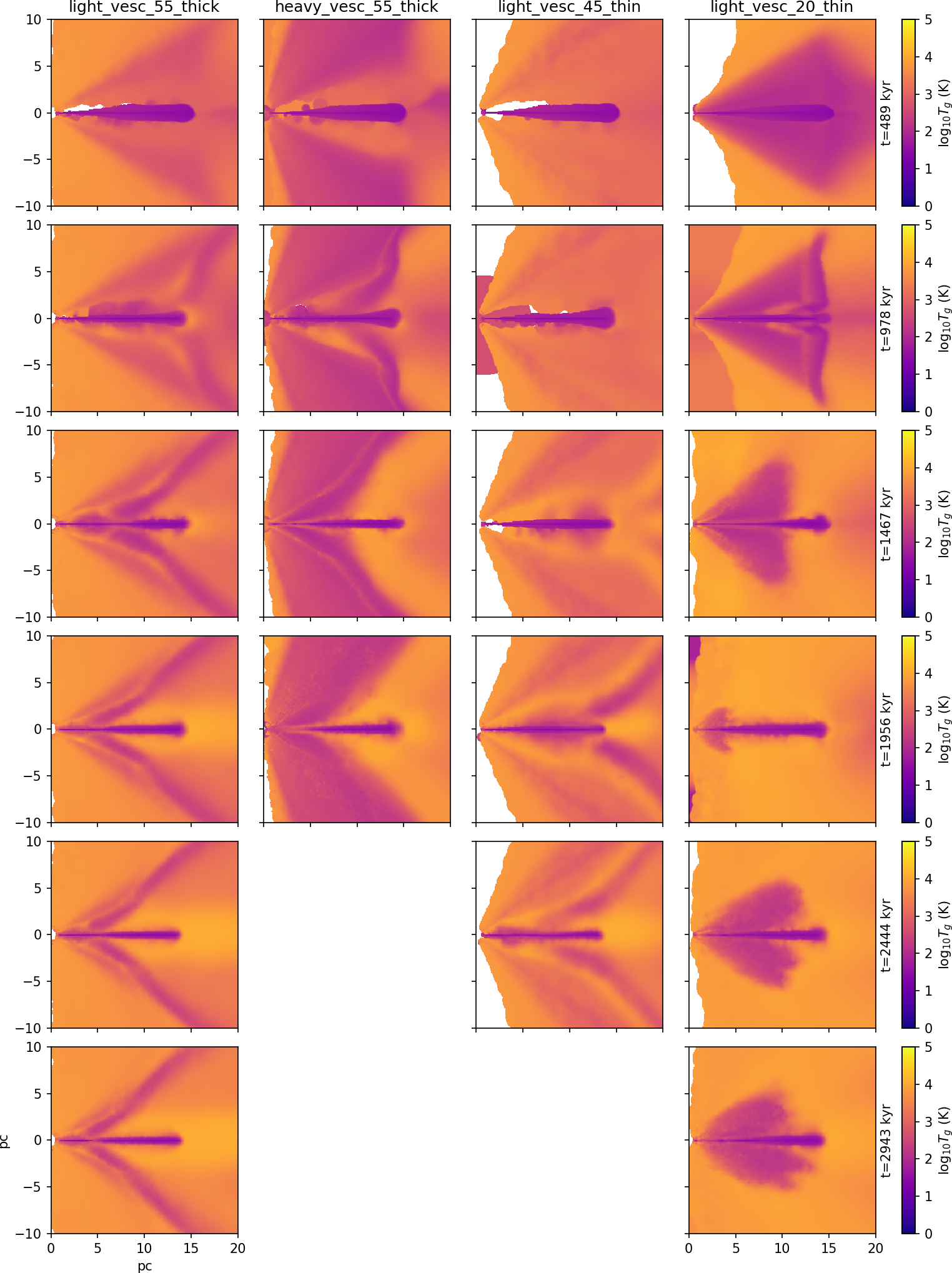}
\end{center}
\caption{\label{evol_Tg}Evolution of azimuthally averaged temperature distribution of four sample full-scale runs }
\end{figure*}

\begin{figure*}
\begin{center}
\includegraphics[width=.95\textwidth]{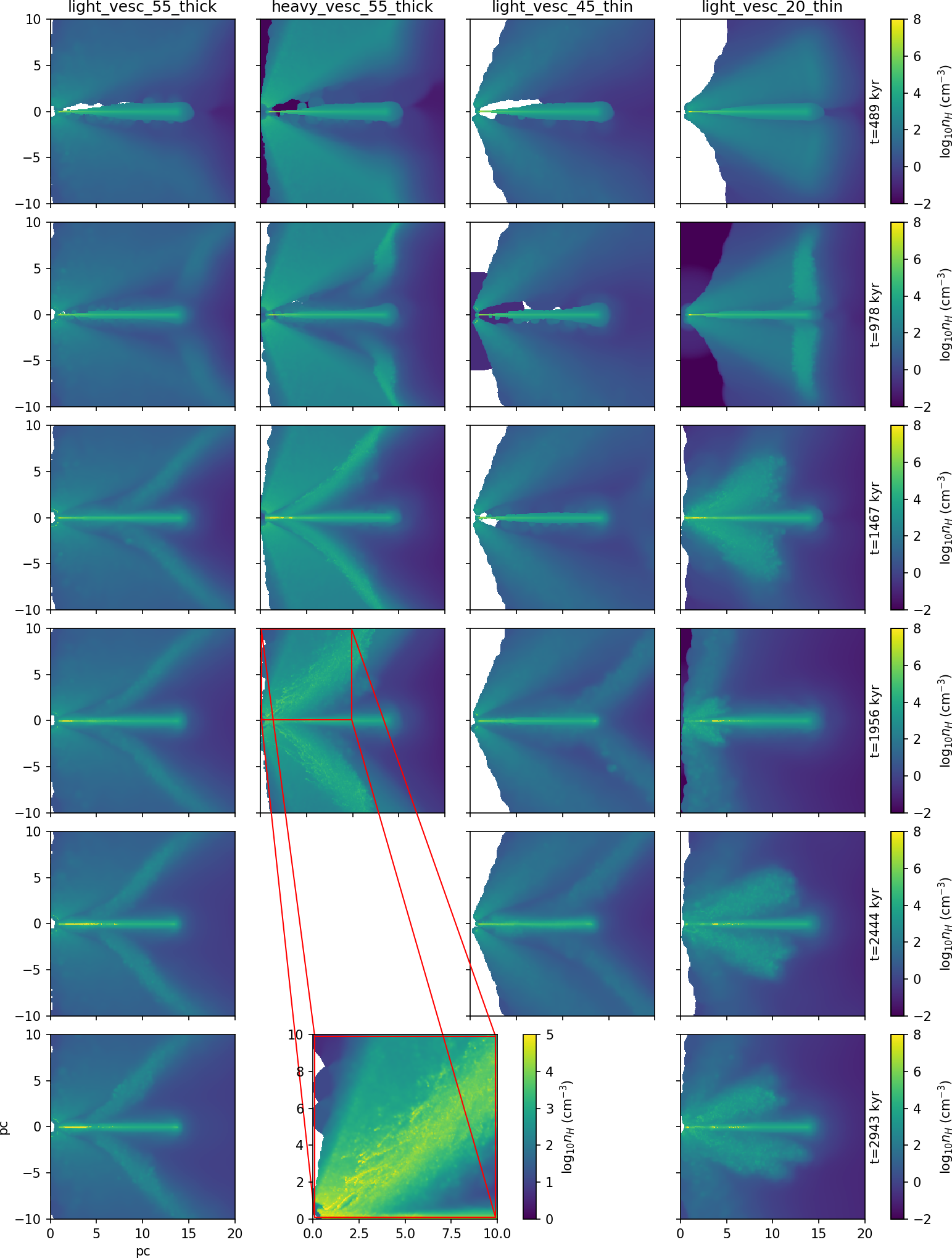}
\end{center}
\caption{\label{evol_nH} Evolution of azimuthally averaged density distribution of four sample full-scale runs. A zoom panel shows the clumpy structure in one run -- note the different color scale).}
\end{figure*}

We have plotted the evolution of the inner region of four sample runs in Figures \ref{evol_v}, \ref{evol_Tg}, and \ref{evol_nH}. These show the azimuthally averaged velocity, temperature, and density of the simulations over several snapshots.

In all runs, the wind initially flows outwards at all angles, except where the disk blocks the outflow. The outflow speed is greatest in the more polar directions. Later, lower angle winds eventually fail and begin to flow back in. A dense front is produced on the interface between successful and failed winds. The system typically settles into a low-density rapid outflow at high inclinations, and a failed wind/inflow at lower inclinations, with a persistent dense biconical front in-between. Generally the temperature follows a fairly simple equation of state, where dense gas is cold and rarefied gas is hot.

This combination of outflow and inflow is a `fountain' structure. Both inflow and outflow are largely radial. This is because the gravitational potential is almost spherically symmetric, as it is dominated by the gravity of the SMBH and the bulge. The gas only makes a secondary contribution to large-scale gravity. Hence, the motion of the wind becomes somewhat Keplerian, in regions where radiation pressure becomes unimportant due to extinction, or due to inclination effects (i.e. because radiation is weaker at more equatorial angles). In the infall phase, the gas is effectively in a highly eccentric elliptical orbit, and so falls back towards the SMBH almost radially. The wind (both inflowing and outflowing phases) also tends to have a low rotation speed, due to conservation of angular momentum over the large distance between the wind production radius (the sublimation radius $r_L=0.015$~pc) and the radii of interest for the fountain ($r\sim1$--$20$ pc). Hence, circular motion is a small contribution, and the wind trajectory is indeed similar to a Keplerian ellipse in a plane.

In some simulations, more complex structure can also form at later times. In light\_vesc\_20\_thin, the failed infalling wind collides with the equatorial outflowing wind to produce a dense turbulent biconical structure. In heavy\_vesc\_55\_thick, the inflowing wind fragments into dense clumps. This clumpiness is seen in the other runs, but is most prominent when the outflow is massive. This appears to be caused by hydrodynamic instabilities at the interface between inflows and outflows (rather than numerical effects from e.g. artificial viscosity), but we will explore the mechanisms and ramifications of this clumpiness in a future paper.

\begin{figure*}
\begin{center}
\includegraphics[width=\textwidth]{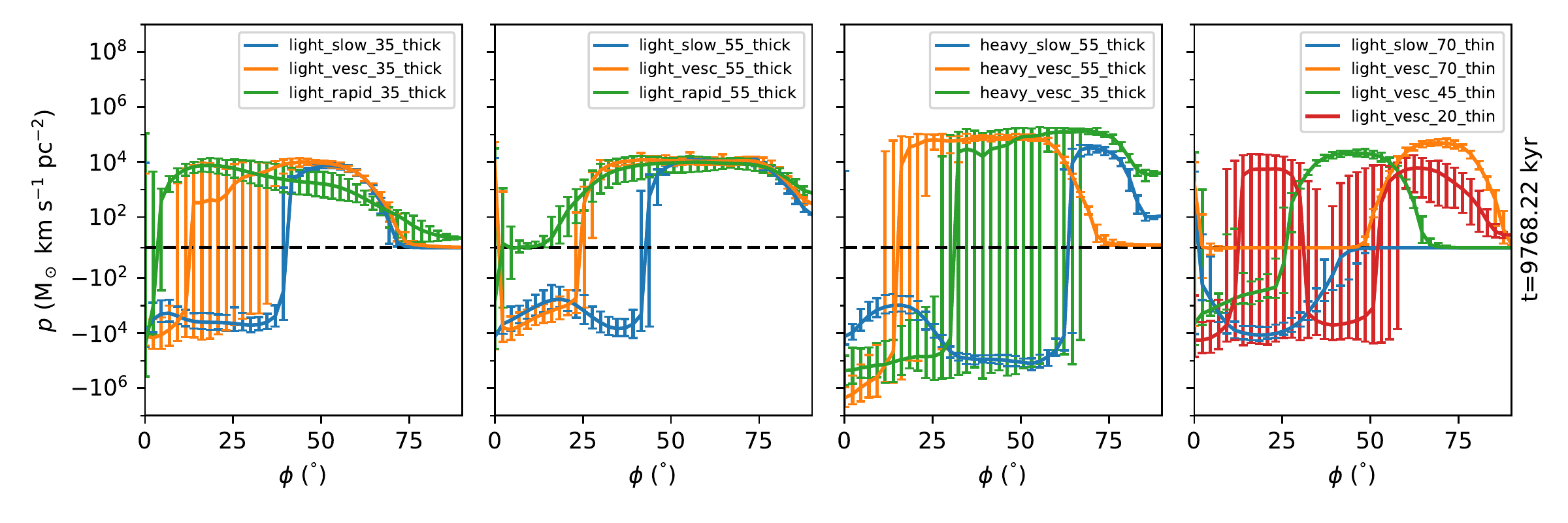}
\end{center}
\caption{\label{mom_angle}Momentum surface density versus inclination at the final snapshot of each of the full scale runs. The vertical axis has a `symlog' scaling. Left two plots: $\dot{M}=0.0378$ M$_\odot$ yr$^{-1}$, $\Delta\phi_w=50^\circ$, $\phi_w=55^\circ$ (first plot) or $\phi_w=35^\circ$ (second plot). Third plot: $\dot{M}=0.378$ M$_\odot$ yr$^{-1}$, $\Delta\phi_w=50^\circ$ runs. Fourth plot: $\Delta\phi_w=20^\circ$}
\end{figure*}

\begin{figure*}
\begin{center}
\includegraphics[width=\textwidth]{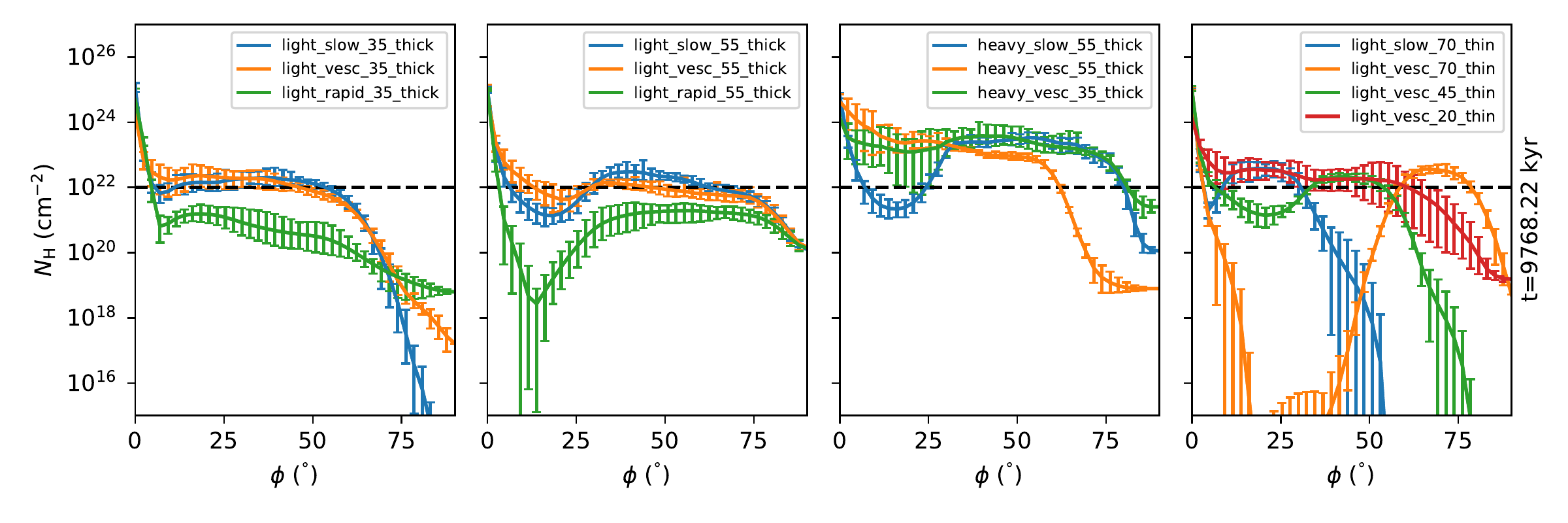}
\end{center}
\caption{\label{NH_angle}Column density versus inclination at the final snapshot of each of the full scale runs. Runs are collected as in Figure~\ref{mom_angle}.}
\end{figure*}

\begin{figure*}
\begin{center}
\includegraphics[width=\textwidth]{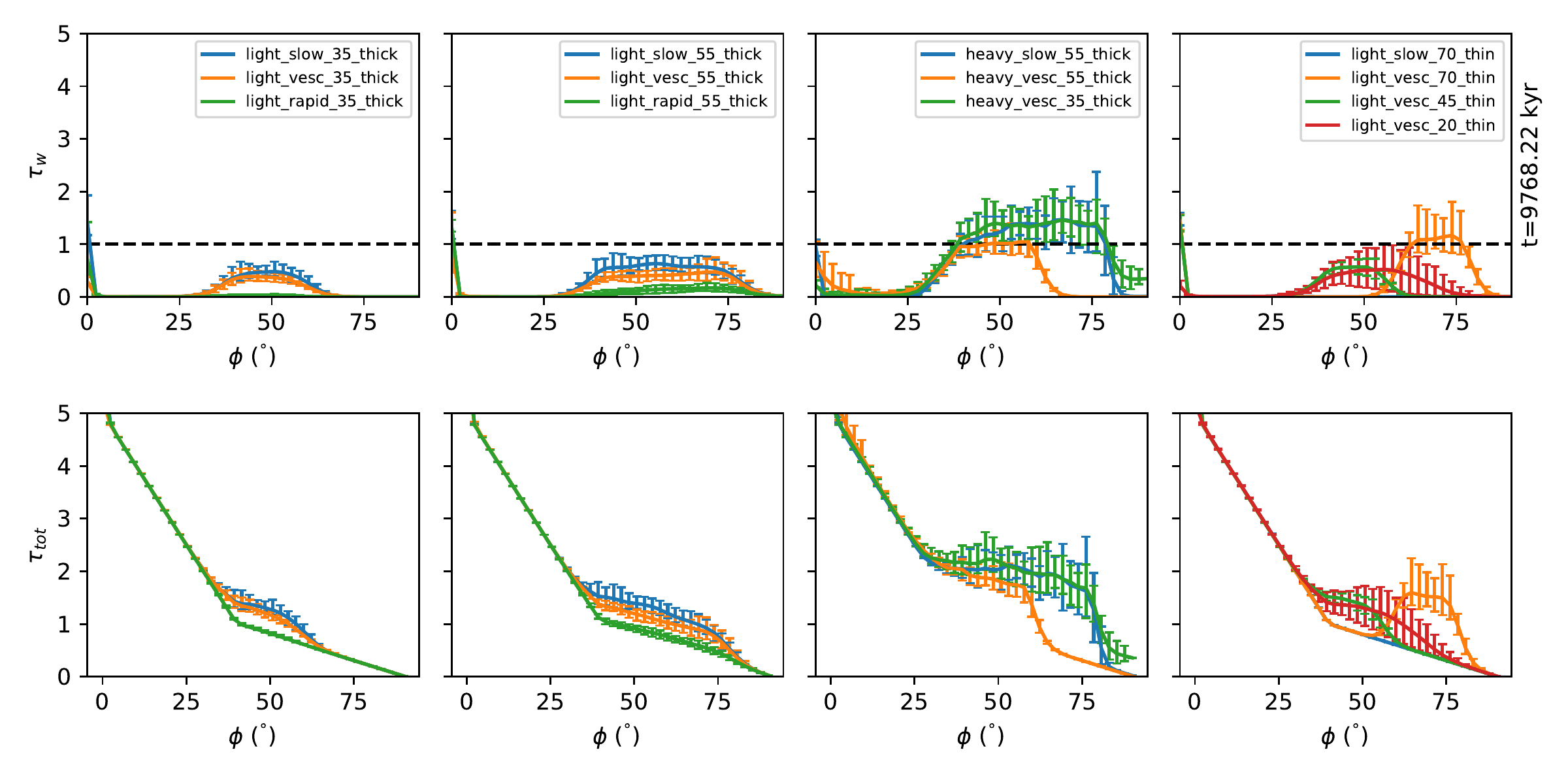}
\end{center}
\caption{\label{tau_angle} Optical depth versus inclination at the final snapshot of each of the full scale runs. Runs are collected as in Figure~\ref{mom_angle}. Top row: optical depths from simulated gust \& dust. Bottom row: optical depths including sub-grid extinction.}
\end{figure*}

We summarize all $14$ of our full-scale runs in Figures~\ref{mom_angle}-\ref{tau_angle}. For each run, we examine the final snapshot of each run, once the wind has settled down into a steady state. We calculate the momentum surface density, mass column density, and optical depth along a grid of rays from the AGN center to infinity. This is a grid of $40$ evenly spaced inclination angles and $40$ evenly spaced azimuthal angles. The azimuthal mean, maximum, and minimum of outflow quantities at each inclination are plotted in these Figures, as lines with errorbars.

The momentum surface density is plotted in Figure~\ref{mom_angle}, where positive $p$ indicates outflow and negative $p$ indicates inflow. Here the fountain structure is clear in almost all runs, with outflows at high angles and inflows at low angles. The only exceptions are the `rapid' runs, where outflows greatly exceed escape velocity and do not fall back in at any angle. All runs have significant outflowing material at very high inclinations, even when winds are launched at low angles. Often there is outflow even at $90^\circ$ -- i.e. `straight up'. In several runs, there is a large range of inclinations where both inflow \textit{and} outflow are present. The system is therefore not azimuthally symmetric, and contains three-dimensional structure. This is the turbulent clumpiness described above.

The contribution of the outflows to the obscuring `torus' is plotted in Figures~\ref{NH_angle}~\&~\ref{tau_angle}. The column densities (Figure~\ref{NH_angle}) can be fairly significant over a large covering fraction, especially for a massive outflow. However, this obscuration is not generally optically thick. The flux-weighted optical depths of the simulated gas are plotted in the top panel of Figure~\ref{tau_angle}, and here $\tau>1$ only for some parts of some runs. In the bottom panels of Figure~\ref{tau_angle} we include the contribution from the sub-grid optical depths in the source, which generally dominates, although the simulated gas adds a significant contribution at high inclinations. So we conclude that most of the obscuration required for the AGN unification scheme is from structure at a sub-parsec scale ($\lesssim70r_{sub}$), and not produced by large-scale outflows.

\subsubsection{Full-scale runs -- mock observations}\label{results_obs}

A key aim of our models is to produce a self-consistent dynamical model to broadly explain the types of morphologies derived from observations and radiation transfer models \citep{2017ApJ...838L..20H,2017MNRAS.472.3854S}. We do not compare our dynamical models directly with specific observations in detail, as this would require a much larger parameter-space search, and a fuller radiation transfer model. This is beyond the scope of this paper. Nevertheless, we can use simple assumptions to make \textit{qualitative} comparisons with observations, and to put our simulations in context.

We use a simple raytracing technique for post-processed radiation transfer. The emissivity of several molecular lines are calculated and tabulated using \textsc{Cloudy} following the same ray-tracing described in Section~\ref{section_existing_model}, and obscuration is traced along the line of sight. The line-of-sight opacity for these narrow lines is given by a simple dusty gas opacity law, extracted from \textsc{Cloudy}. The estimate of infrared emission $F_{IR}$ for a particle is simply proportional to $T_d^4$, where $T_d$ is the dust temperature, as previously tabulated using \textsc{Cloudy}. We set the broadband infrared opacity of the dusty gas to $65.2$ cm$^{2}$ g$^{-1}$. This is the opacity of a dust-gas mix where the sole source of extinction is from perfectly absorbing spherical dust particles with a radius of $0.2$~$\mu$m, an internal density of $6\times10^3$~g\,cm$^{-3}$, and a total dust mass fraction of $1\%$.

We emphasize that a `critical density' is not sufficient to completely characterize the behavior of a line, without further information about the radiation field. In our simulations, the line emissivities are a function of gas density, gas temperature, unextinguished AGN intensity, and flux-weighted optical depth to the AGN. The dust temperature also depends on these properties, although only weakly on density. Hence we find we can largely characterize the emissivity as a function of two variables -- the gas density and the dust temperature. The dust temperature acts sort of like a low-energy `ionization parameter', collecting the effects of the radiation field and gas temperature into a single parameter. In particular, we find three different `regimes' of dust temperature. 

Only a small number of particles close to the AGN engine contain hot dust ($T_d>100$ K), and we find these line emissivities all decrease rapidly with increasing dust temperature, representing a more intense radiation field destroying molecules (note that typically $T_g>T_d$). There is also a somewhat ragged increase of emissivity with density for all lines. For the `cold'  ($T_d<15$ K) and `warm' ($15<T_d<100$ K) dust, we find that the emissivities are generally constant at low densities, then start rising at $n_H\sim10$~cm$^{-3}$, before reaching a peak at some point. Examining a sample run in detail, we found two peaks at different emissivities at high $n_H$, which we found corresponded to gas populations with two different dust temperatures -- the emissivities from gas with cold dust were typically lower. The `cold' dust phase represents outflowing gas far from AGN illumination.

\begin{figure}
\begin{center}
\includegraphics[width=\columnwidth]{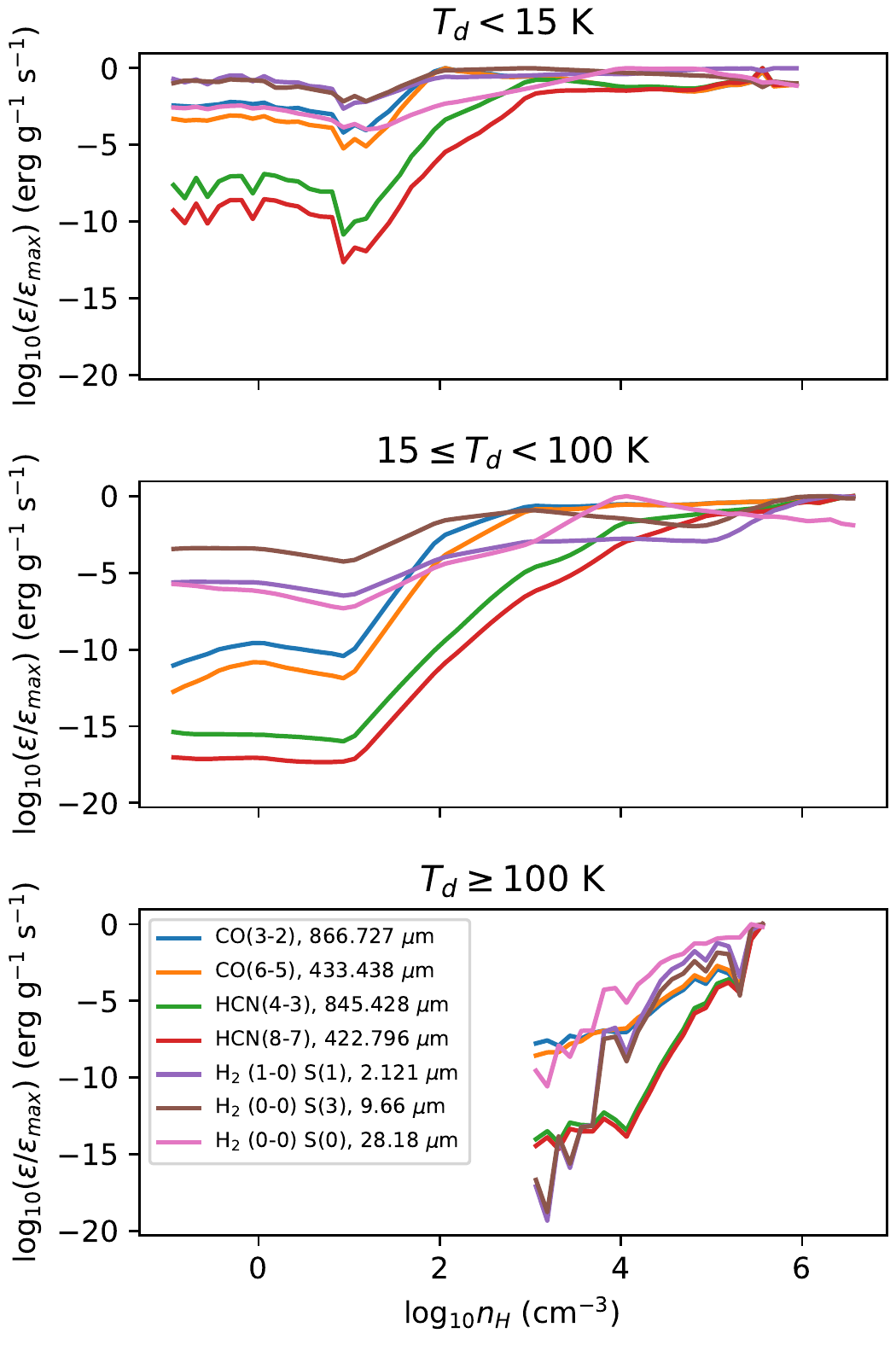}
\end{center}
\caption{\label{line_explain}
Normalized median emissivities of simulation heavy\_vesc\_55\_thick at $t=2$ Myr in density and dust temperature bins.}
\end{figure}

In Figure~\ref{line_explain} we plot the normalized median line emissivity for various values of $n_H$ for these three dust temperature regimes, in a snapshot of run heavy\_vesc\_55\_thick at $t=2$ Myr. The `warm' dust regime contains the bulk of the mass and emission, and so we concentrate on the middle panel here, although the `cool' dust regime shows similar behavior in most cases. The two HCN lines follow similar paths, both steadily increasing with $n_H$ up to the maximum densities found in the simulation. These lines highlight the densest regions of gas, with HCN(8-7) emphasizing slightly denser gas than HCN(4-3). Both CO(3-2) and CO(6-5) increase from $n_H=10$~cm$^{-3}$ before flattening out at around $n_H=10^3$~cm$^{-3}$. These emphasize a mix of moderate-to-high density gas, from around $n_H=10^2$~cm$^{-3}$ upwards. Both lines produce very similar emission. The \MH lines have a shallower slope with respect to density, and therefore emphasize an even broader range of densities, including low density gas. Here there is a notable difference between the `cool' and `warm' dust phases. In both phases, \MH (0-0) S(0) is biased towards higher densities than \MH (0-0) S(3), but \MH (1-0) S(1) is more biased towards higher densities in the `warm' phase, and flatter in the `cool' phase.

\subsubsection{Full-scale runs -- mock morphology observations}\label{results_obs}

\begin{figure*}
\begin{center}
\includegraphics[width=\textwidth]{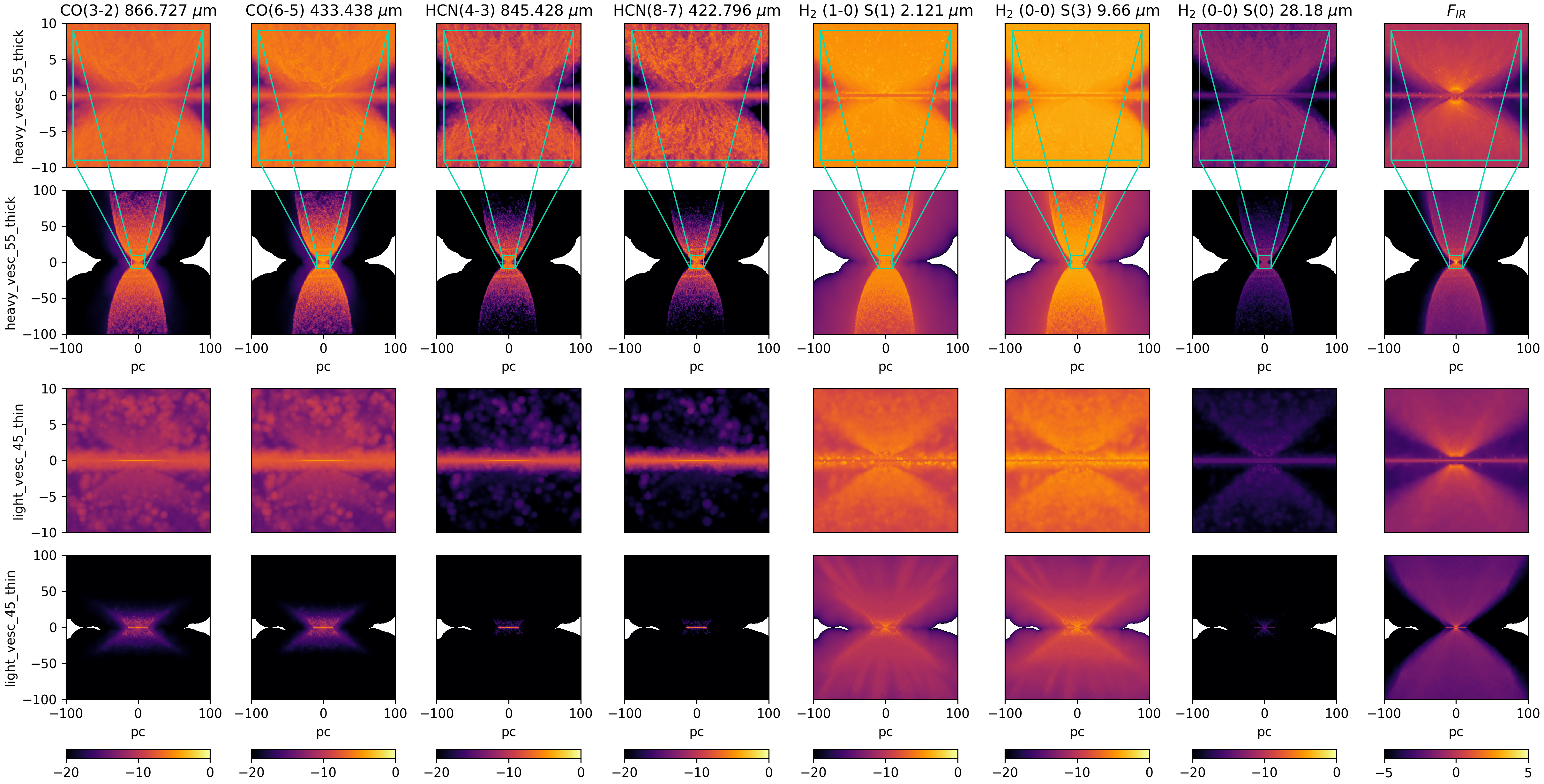}
\end{center}
\caption{\label{line_summary}
Emission for various lines, and broad thermal emission from dust ($F_{IR}$), at $10$ pc and $100$ pc scales, for runs heavy\_vesc\_55\_thick and light\_vesc\_45\_thin at $t=2$ Myr.
}
\end{figure*}

To demonstrate the general observed morphology, we produce imagery of two sample runs, run light\_vesc\_45\_thin and run heavy\_vesc\_55\_thick at $t=2$ Myr in Figure~\ref{line_summary}. The images are given at a $100$ pc scale, and a $10$ pc scale. Here it is clear how different molecular lines highlight different components, and also that the two models produce noticeably different images. As expected, we find that our two CO lines produce very similar images to each other, as do our two HCN lines, and our three \MH lines, and so we will discuss each set of lines as a group. \MH (0-0) S(0) is dimmer than the other \MH rotational lines, but highlights gas in fairly similar states.

Generally, the CO and HCN lines therefore highlight the dense disk, and the high-density portions of the outflow. The HCN lines emphasize higher densities than the CO lines. So for the massive outflow of heavy\_vesc\_55\_thick, the outflow is visible in both CO and HCN, but appears more extended in CO. But for the less massive outflow of light\_vesc\_45\_thin, the outflow is only visible in CO, and only the disk is visible in the HCN lines. Substructure is also visible in the outflow, in the form of sub-parsec scale clumps and filaments. The dust is optically thin at these long wavelengths, and so the disk is brightest through the large column densities in the centre -- i.e. it is limb-darkened.

The \MH lines emphasize the lower density outflow much more than the CO and HCN lines. This means the outflow is visible even in light\_vesc\_45\_thin. Here the thin-walled outflow cone is limb-brightened, producing an x-shaped structure. The thicker-walled outflow of heavy\_vesc\_55\_thick does not show this x-shape -- a limb brightened x-shape is an indicator of a thin-walled cone. Dust is fairly opaque to these shorter wavelengths, as required by AGN unification. The disk is therefore self-obscuring, and only visible from edge-on as a shadow.

In light\_vesc\_45\_thin, radial beams of \MH emission are also visible. These are caused by `windows' in obscuration near the AGN engine. Examining an animation of this run, we see these radial beams flicker rapidly -- the `windows' are caused by short-scale variations in `weather' near the centre of the simulation. In heavy\_vesc\_55\_thick, the walls of the outflow cone are too thick for these beams to be visible.

The dust temperature and therefore $F_{IR}$ depends strongly on the incident radiation field. The outflows are visible as an extended hourglass, but the peak brightness is at the base of the wind, where the dust is heated the most.

This hourglass structure is not caused by any radiative process accelerating the gas vertically -- radiation pressure is strictly radial in these simulations. After $\sim10$s of pc, the gas is moving ballistically on hyperbolic orbits, producing these vertical hyperbolic-like hourglass structures, provided the outflow is somewhat vertical before reaching the ballistic phase. Note that, although the orbits are hyperbolic, this structure is sometimes referred to as a `parabolic' hollow cone.

\begin{figure}
\begin{center}
\includegraphics[width=\columnwidth]{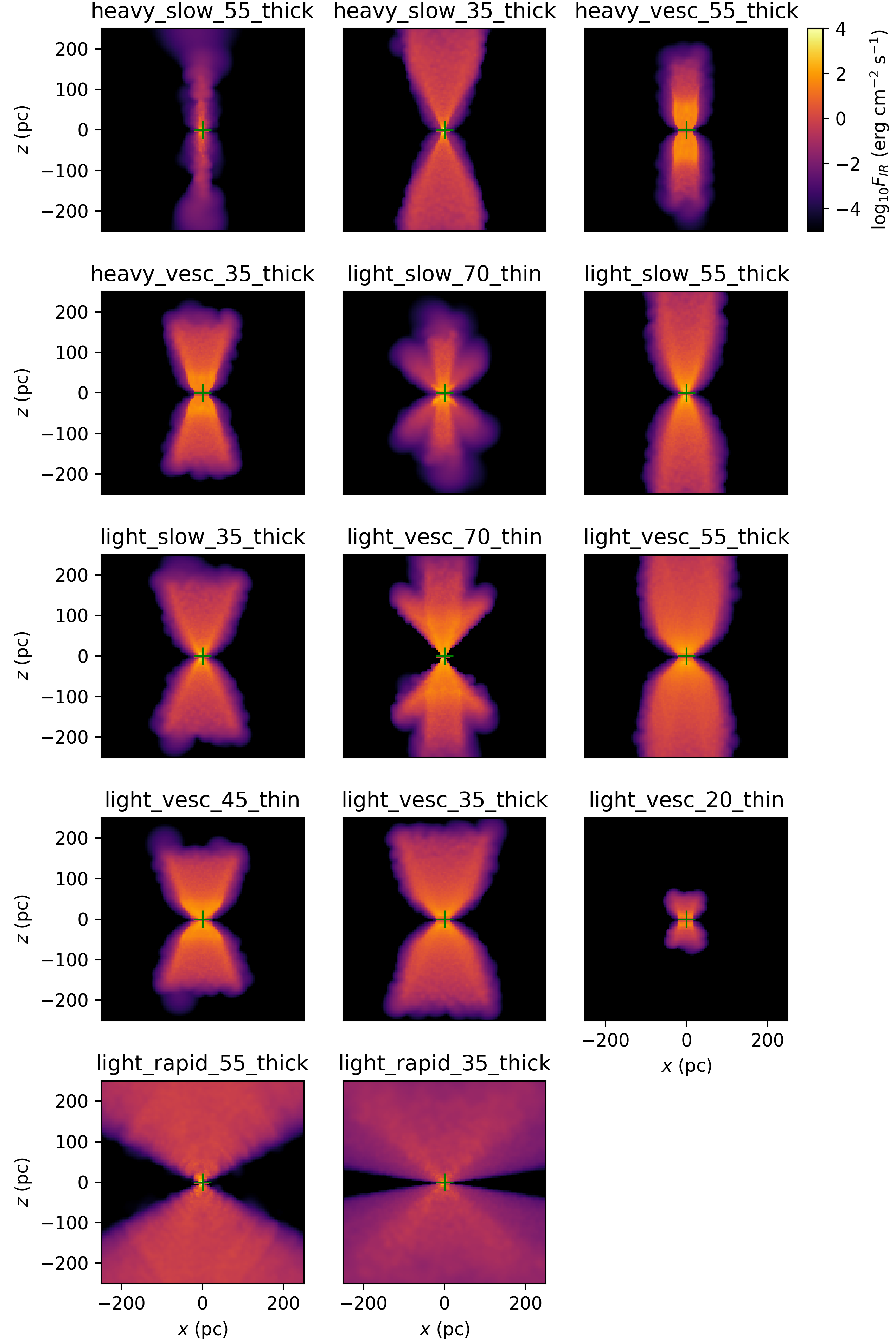}
\end{center}
\caption{\label{ir_outflow} Broad thermal emission from dust ($F_{IR}$), for all full-scale runs, at $t\approx1$~Myr.}
\end{figure}

In heavy\_vesc\_55\_thick, the injected wind is slightly polar by design -- i.e. it is assumed that some small-scale process is preferentially launching a somewhat polar wind (section~\ref{section_shortruns_results}). But in light\_vesc\_45\_thin, the wind is launched at $45^\circ$, and evolves into a more polar wind through the anisotropic radiation pressure of the AGN accretion disk. We see similar results in the other runs. In Figure~\ref{ir_outflow}, we plot $F_{IR}$ on a $500$~pc scale for all of our full-scale runs. In almost all cases, polar extended IR emission is observed, regardless of the wind launching angle. These polar extended structures appear to have a robust mechanism for their formation, and are not produced by fine-tuning the input parameters. The AGN flux and radiation pressure is stronger in the polar direction, and so almost any distribution of dusty gas will tend to be pushed in the polar direction, and be illuminated by the AGN. The only exceptions are the `rapid' runs, where the wind speed is so high that the effects of gravity and radiation pressure are almost negligible.

\begin{figure*}
\begin{center}
\includegraphics[width=\textwidth]{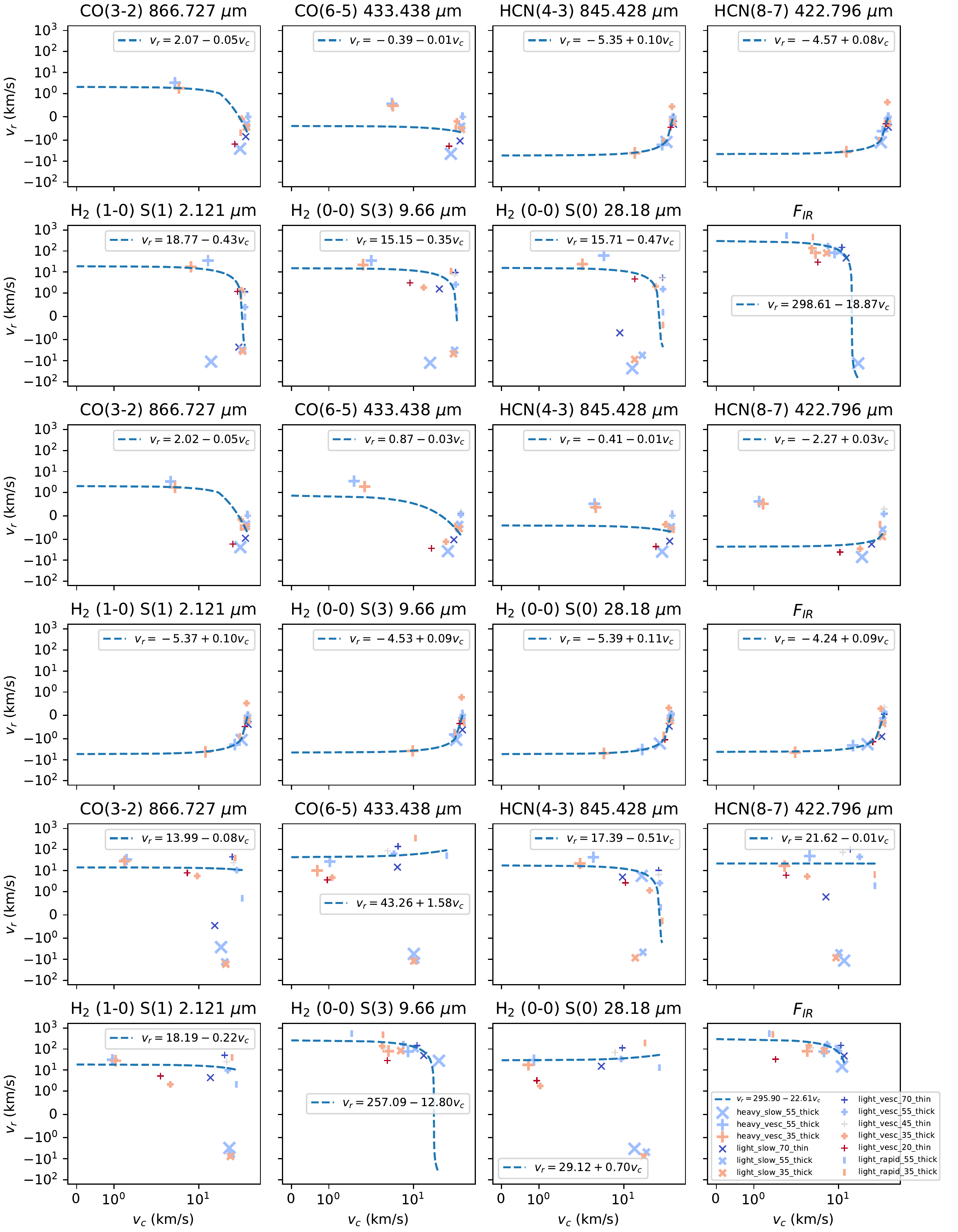}
\end{center}
\caption{\label{lum_v} Emission-weighted radial and circular velocities for different emission lines, for all runs at $\approx1$~Myr. Top: unextinguished emission. Middle: extinguished emission, face-on view. Bottom: extinguished emission, edge-on view}
\end{figure*}

The kinematic structures captured by the different lines are shown more systematically for all runs in Figure~\ref{lum_v}. For the top two rows, we calculate the emissivity weighted average radial and circular velocities for each molecular line (and broadband $F_{IR}$), for all runs at $t=1$ Myr. Some of the features are better at distinguishing outflow features than others. The CO and HCN lines emphasize dense gas, and mostly detect inflowing and rotating material, except for the heaviest outflows in the CO lines. The \MH lines capture more outflowing gas and less rotation. There is a correlation between radial velocity and circular velocity in \MH -- outflowing material has a lower rotation speed. The infrared thermal emission selects the wind most effectively, showing high outflow velocities and low circular velocities.

\begin{figure*}
\begin{center}
\includegraphics[width=\textwidth]{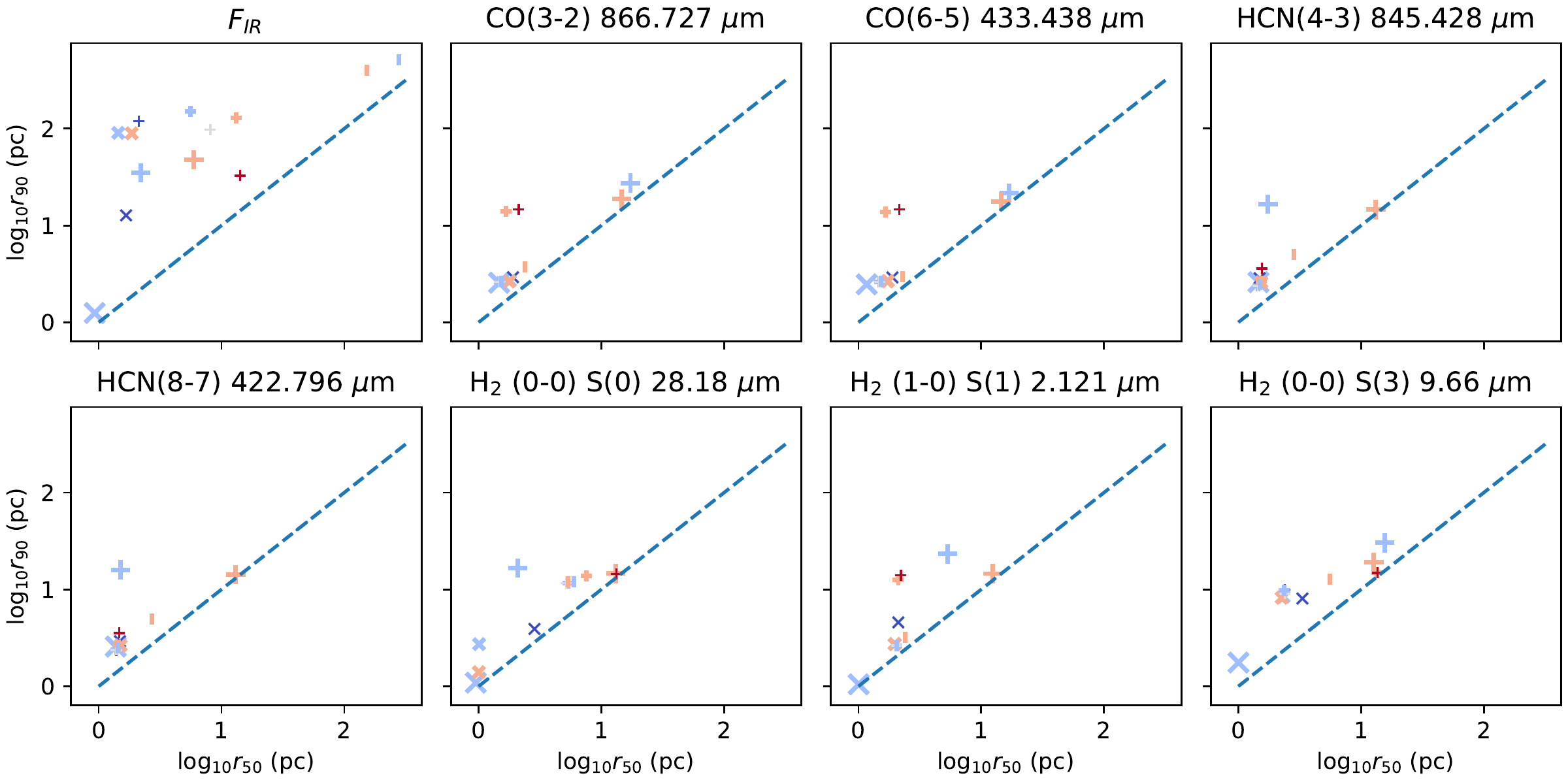}
\includegraphics[width=\textwidth]{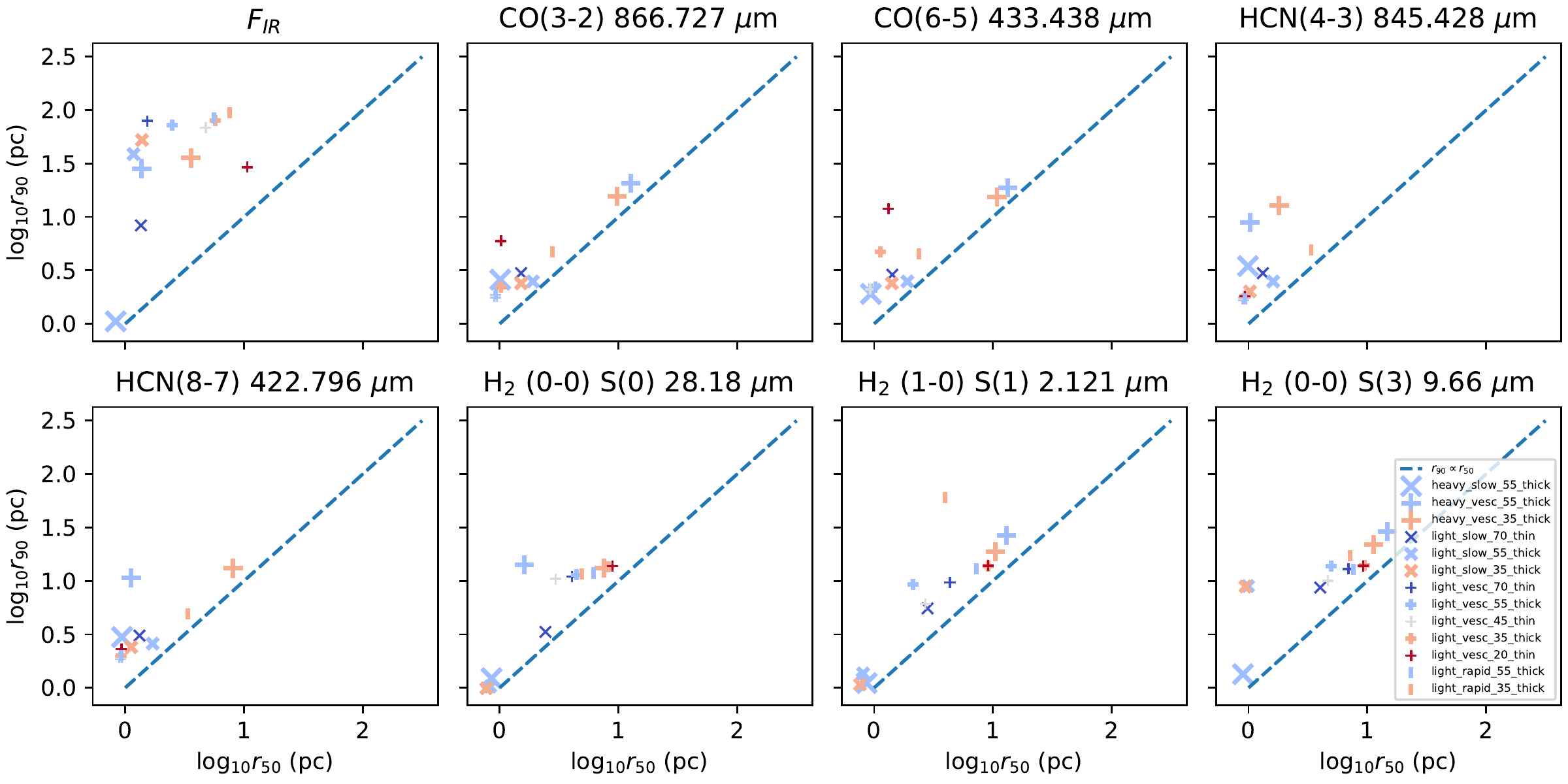}
\end{center}
\caption{\label{lum_r} Half-light and $90\%$-light radii for different emission lines, for all runs at $\approx1$~Myr.  Top: unextinguished emission. Bottom: extinguished emission, edge-on view
}
\end{figure*}

We include the effects of line-of-sight extinction in the bottom four rows of Figure~\ref{lum_v}, from a face-on view (rows~3~\&~4) and an edge-on view (rows~5~\&~6). The face-on view of CO and HCN lines still mostly emphasizes the disk and outflow, because the dust is mostly transparent in these wavelengths. However, in the heaviest outflows the disk and inflow are less visible, and the outflow is emphasized. The effect of extinction is stronger in the \MH lines, increasing towards shorter wavelengths. The edge-on view greatly emphasizes the outflowing gas in all lines for most runs -- the large column density of the disk and the thick inflow cause them to become optically thick, and their emission is less visible.

In Figure~\ref{lum_r}, we calculate the radii within which $50\%$ and $90\%$ of the luminosity of each line (or $F_{IR}$) is produced, labelled as $r_{50}$ and $r_{90}$. Similarly to Figure~\ref{lum_v}, in the top two rows, these are calculated without extinction -- i.e. this is the physical radius from which this emission is produced -- and in the bottom two rows, these are calculated with line-of-sight emission from an edge-on view. $r_{50}$ and $r_{90}$ are another measure of the distribution and concentration of the emission, and the plot demonstrates how the distribution varies based on run properties. We also plot a line of $r_{50}\propto r_{90}$. Luminosity distributions that vary only in scale and not in concentration will lie parallel to this line.

The HCN lines particularly emphasize the dense gas of the disk, and therefore have small uncorrelated $r_{50}$ and $r_{90}$, except for the more massive outflows. Here we can indeed distinguish the wind angle for the massive outflows: the more vertical wind produces a dense fountain that concentrates $r_{50}$, while the more horizontal wind distributes HCN emission more evenly.

The \MH lines capture the outflow most effectively, and show the strongest correlation between $r_{50}$ and $r_{90}$. The strongest outflows have the highest $r_{50}$ and $r_{90}$.

The CO lines lie in-between, showing several runs with low $r_{50}$ and $r_{90}$, but with large $r_{50}$ and $r_{90}$ for the most massive outflows. Here the angle of the outflow can not be easily distinguished between those two runs based on the CO distribution alone.

Overall, we see that the concentration of emission in a particular line is not a strong indicator of the wind structure. Many lines emphasize the disk and do not capture the outflow well at all, and when the lines do capture the outflow, the different winds simply produce emission of different extents.

\subsubsection{Full-scale runs -- mock velocity observations}

\begin{figure*}
\begin{center}
\includegraphics[width=\textwidth]{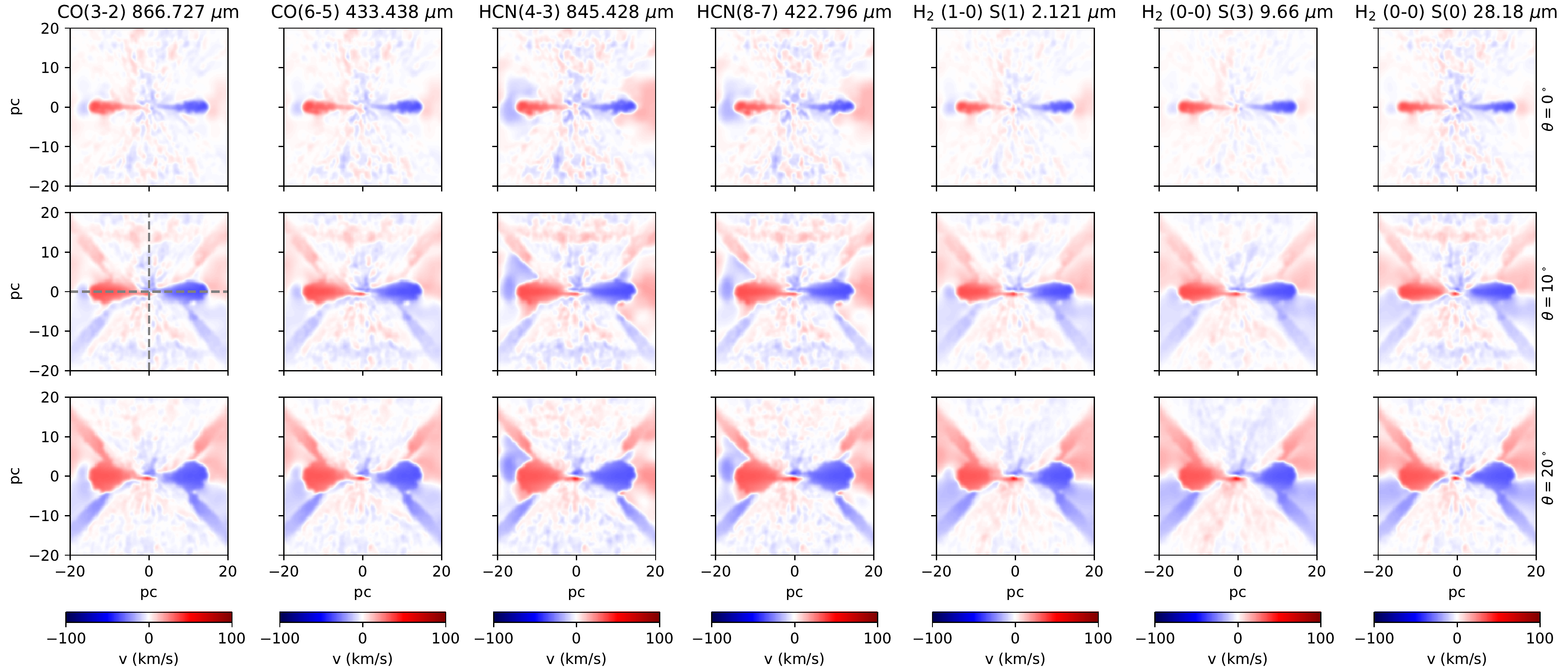}
\end{center}
\caption{\label{vmap1}
Mean los velocities at $t=2$ Myr for run heavy\_vesc\_55\_thick. The dotted lines indicate the vertical and horizontal slits used for Figure~\ref{v_pos_maps}.}
\end{figure*}

\begin{figure*}
\begin{center}
\includegraphics[width=\textwidth]{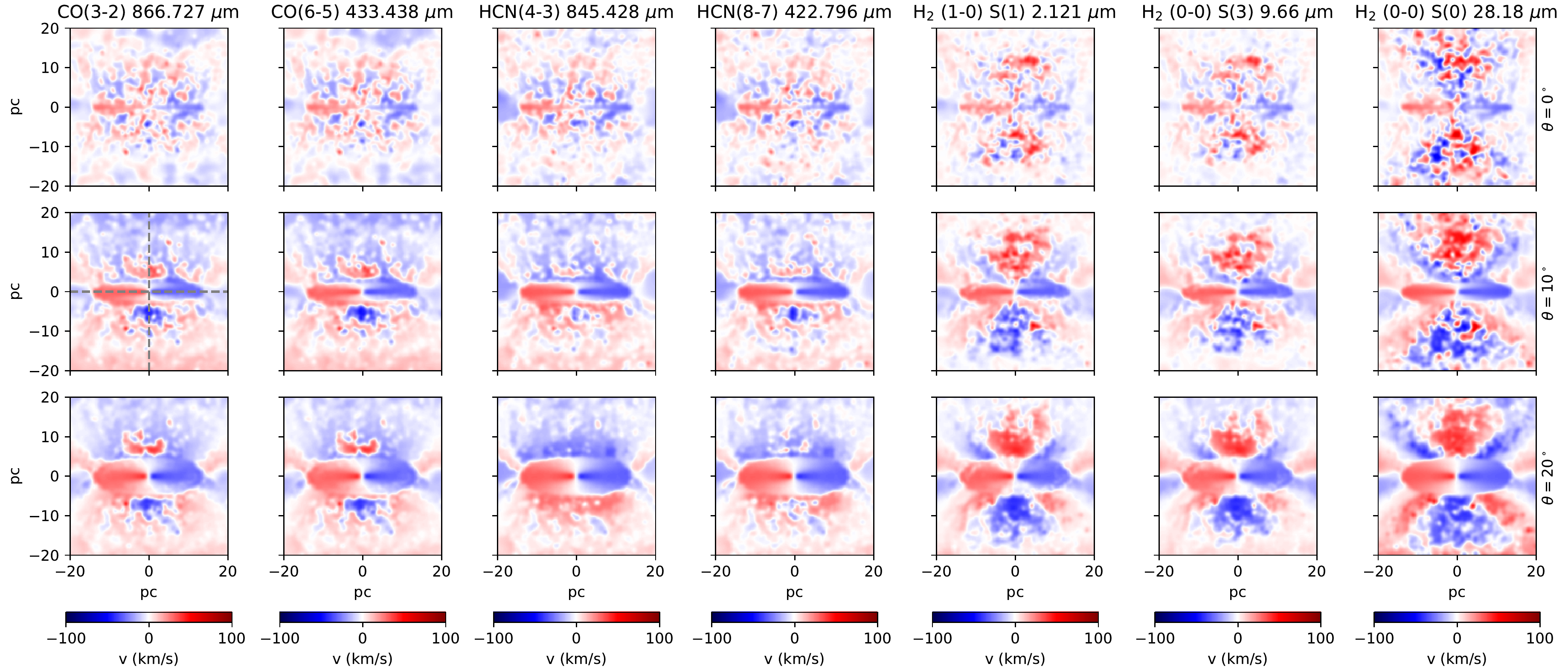}
\end{center}
\caption{\label{vmap2}
Mean los velocities at $t=2$ Myr for run light\_vesc\_45\_thin. The dotted lines indicate the vertical and horizontal slits used for Figure~\ref{v_pos_maps}.}
\end{figure*}

We produce mean line-of-sight velocity maps for our seven molecular emission lines at $t=2$ Myr from near edge-on viewing angles, using line-of-sight extinction as in Section~\ref{results_obs}. We plot the results for heavy\_vesc\_55\_thick in Figure~\ref{vmap1}, and for light\_vesc\_45\_thin in Figure~\ref{vmap2}. These velocity maps are convolved with a $0.4$~pc radius Gaussian, to represent a finite beam size. This is about $6$~mas at the distance of NGC 1068, about $10\times$ finer than recent ALMA observations \citep{2019A&A...632A..61G}, and similar to the latest GRAVITY observations \citep{2020A&A...634A...1G}.

The disk rotation is clearly visible, but is partially hidden by the signal of the outflowing gas, particularly in heavy\_vesc\_55\_thick. These images include several other features that could be misinterpreted as counter-rotation or non-aligned rotation. In heavy\_vesc\_55\_thick (Figure~\ref{vmap1}), there is a strong outflow signal in the center where outflows are generated, appearing as a sharp gradient between blue-shifted and red-shifted emission for several lines. Beyond the central parsec, the clumpy mix of outflow and inflow hides the outflow signal, except at the edge of the outflow, where an `x' shape appears. This could be misinterpreted as a pair of crossed discs, at $\sim45^\circ$ to the AGN plane.

The weaker outflow does not hide the disk of light\_vesc\_45\_thin (Figure~\ref{vmap2}), but the more uniform outflow produces a clearer signal, especially at higher viewing angles and the \MH lines. This simple model of a disk plus outflow/fountain again produces a number of velocity gradients in different directions, most of which are produced by the radial flow rather than rotation.

\begin{figure*}
\begin{center}
\includegraphics[width=\textwidth]{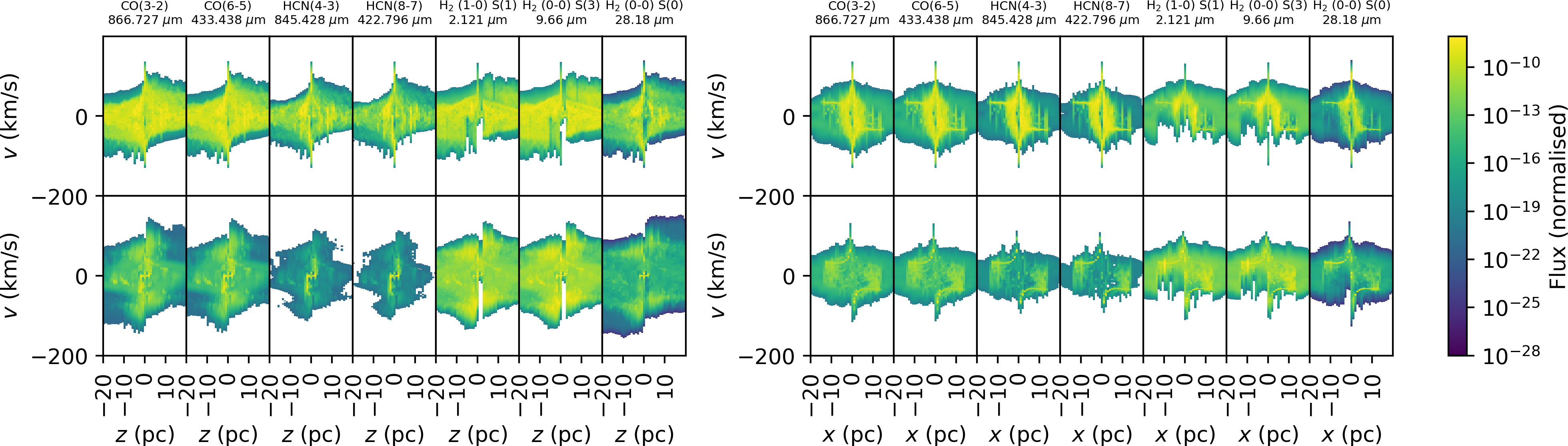}
\end{center}
\caption{\label{v_pos_maps}
Position-velocity diagrams for runs heavy\_vesc\_55\_thick (top row) and light\_vesc\_45\_thin (bottom row), in a vertical slit (left) and horizontal slit (right), from a viewing angle of $10^\circ$.}
\end{figure*}

However, mean line-of-sight velocity maps can be misleading, as inflow superimposed on outflow can appear as near-stationary gas. So in Figure~\ref{v_pos_maps} we plot the full position-velocity diagrams for a vertical and horizontal slice at a viewing angle of 10$^\circ$ for both sample simulations and all lines. For clarity, these axes are plotted on the center-left plots of Figures~\ref{vmap1}~\&~\ref{vmap2}.

Disk rotation is not visible in the vertical slit, as this is the axis where the disk velocity is tangential to the line of sight. Instead, outflows and inflows dominate the position-velocity diagram. The fountain dynamics produce a wide range of velocities in every ray, and because the viewing angle is close to edge-on, there is only a weak velocity slope with $z$. The thin-walled biconical outflows of light\_vesc\_45\_thin are nevertheless visible as two peaks at around $\pm80$ km/s in the inner $\sim10$ pc, in the CO and \MH lines. These outflows are also present in heavy\_vesc\_55\_thick, as the upper and lower edge of the $v$ distribution, but strong fountain inflows of heavy\_vesc\_55\_thick `fill in' the intermediate velocities. This produces a broader peak centred on $v=0$, except at $z=0$, where inflow and outflow velocities reach a peak, following a near-Keplerian infall curve (note that ballistic radial infall/outflow has the same power-law index as circular rotation).

Dust is optically thin in the wavelengths of the HCN and CO lines we have selected, and we see a superposition of outflow and inflow along each line of sight. In our hollow cone outflow+fountain structure, gas flows outwards at the inner edge of the hollow cone, and falls inwards along the outer edge. Each line of sight through the bicone sees the following velocity phases, in order from nearest to furthest:
\begin{enumerate}
\item red-shifted inflow along outer edge of biconical flow, on near-side of AGN
\item blue-shifted outflow along inner edge of biconical flow, on near-side of AGN
\item red-shifted outflow along inner edge of biconical flow, on far-side of AGN
\item blue-shifted intflow along outer edge of biconical flow, on far-side of AGN
\end{enumerate}
Hence through each line of sight, we see outflow and inflow on the near side, as well as inflow and outflow on the far side.

Extinction does have some effect on the p--v diagrams for the \MH lines, but only in the case of heavy\_vesc\_55\_thick where the outflow is particularly thick. Here the blue-shifted emission is blocked, at least for the shorter wavelength \MH (0-0) S(3) and \MH (1-0) S(1) lines. The near-side inflow on the outer edge of the hollow cone is sufficiently optically thick in these wavelengths to block the emission of the near-side outflow on the inner edge of the hollow cone. That is, component (1) of the list above is optically thick, and components 2-4 are not visible.

The horizontal slit shows a much stronger Keplerian disk rotation signal, and the non-circular flows are less dramatic. Here, the wind is only visible when it is quite close to the disk surface -- i.e. very equatorial outflows, or the inflow phase of a fountain returning along a near-equatorial path. There are no strong equatorial outflows. The inflow signal is particularly strong for heavy\_vesc\_55\_thick, producing a `catseye' shape. Here, peak line-of-sight velocities become larger as $x\rightarrow0$, from Keplerian infall. At $x\sim0$, the line of sight velocity for an equatorial flow is close to its true radial velocity relative to the AGN. Hence there is a gap at low velocities around $x\sim0$ -- gas near the SMBH must be moving.

Here, the velocity phases we see along a line of sight are, from nearest to furthest:
\begin{enumerate}
\item red-shifted inflow along far outer edge of biconical flow (near-equatorial), on near-side of AGN
\item disk rotation
\item blue-shifted inflow along far outer edge of biconical flow (near-equatorial), on far-side of AGN
\end{enumerate}
Extinction also has a similar effect in the \MH line p--v diagrams as above. The near-side inflow, and the disk, block much of the emission from the far-side inflow, particularly in the \MH (0-0) S(3) and \MH (1-0) S(1) lines.

The simulations show complex velocity structures. We caution against an oversimplified interpretation of velocity maps and p--v diagrams. As we have found, each line of sight can intersect inflow, outflow, and disk rotation. In unresolved observations, this combination of winds and rotation could appear as a single broad velocity distribution, but this should not be interpreted as entirely due to either rotation or outflows alone.

\subsubsection{Higher Eddington luminosity run}\label{section_highedd}

\begin{figure}
\begin{center}
\includegraphics[width=\columnwidth]{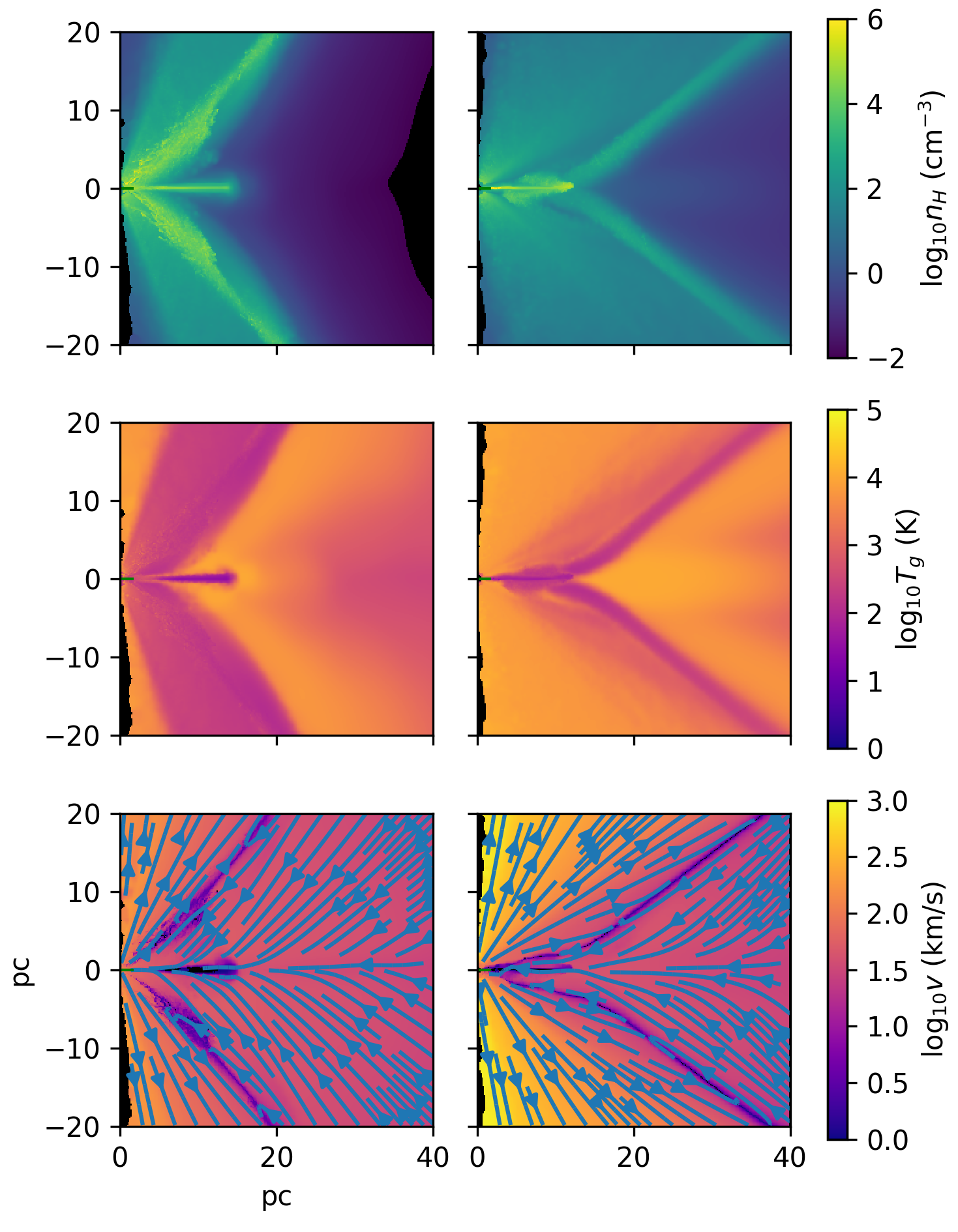}
\end{center}
\caption{\label{highedd}
Azimuthally averaged densities, temperatures, and velocities, to show the effects of higher Eddington factor. Left column: heavy\_vesc\_55\_thick ($\gamma_{Edd}=0.01$). Right column: Identical run except with $\gamma_{Edd}=0.08$.}
\end{figure}

While it was not our intent to perform a full parameter study of SMBH mass and Eddington factor, or to focus on modelling any particular observed AGN, our Eddington factor ($\gamma_{Edd}=0.01$) is significantly lower than is typically observed for AGNs of this mass ($10^6$~M$_\odot$). Hence we have performed a single run at a higher Eddington factor ($\gamma_{Edd}=0.08$) as a sanity check. Other than Eddington factor, this simulation has identical parameters to heavy\_vesc\_55\_thick.

We plot a comparison between heavy\_vesc\_55\_thick and our high Eddington factor run at $t\approx2$ Myr in Figure~\ref{highedd}. Here, the results are as we might expect intuitively and from previous work \citep[e.g.][]{2015ApJ...812...82W}. At a higher Eddington factor, the wind is faster, the biconical structure has a wider opening angle, and the inflow impacts the disk at a greater radius. We still see the fountain structure of vertical outflow and equatorial inflow with a cool dense front in-between. The winds are of course qualitatively different, but there does not appear to be a transition in quantitative properties.

\section{Discussion}\label{section_discussion}

\subsection{Comparison with resolved velocity maps}

We can qualitatively compare our p-v maps with those of NGC 1068 \citep{2019A&A...632A..61G,2019ApJ...884L..28I}, NGC 3227 \citep{2019A&A...628A..65A}, and Circinus \citep{2018ApJ...867...48I}. Our model parameters are closest to those of Circinus (although at a lower $\gamma_{edd}$), and a similar basic structure is seen -- a central `blob', and a sharp peak of red and blue shift in the center. In our models, this structure is produced by a disk and biconical outflow, where the outer part of the cone is a failed wind -- i.e. there is no real counter-rotating component. The p-v diagrams of NGC 1068 and NGC 3227 show a similar structure, but lack the blue-shifted peak in the middle. This may be due to opacity or continuum contributions that become more significant at higher masses, and we will examine these details in an upcoming paper.

\subsection{Comparison with radiation transfer models}

We have not produced a detailed spectrum of our model for comparison with observations. An accurate spectrum requires a more complex radiation transfer scheme, and the lines can be sensitive to sub-grid features such as sub-grid clump optical depth \citep{2017MNRAS.464.2139A}, and the detailed grain population \citep{1993ApJ...402..441L}. We would be testing these post-processing choices at least as much as our dynamical model.

However, our model provides a dynamical explanation for the wind geometries invoked in pure RT models to fit observed spectra \citep{2017ApJ...838L..20H}. These find that a dusty wind can explain the apparent anisotropy of mid-IR emission (as the wind is visible at any inclinations), and can explain the polar extended IR emission. A hollow-cone plus disk geometry, as expected from a wind, can reproduce observed IR spectra \citep{2017MNRAS.472.3854S}. However, on smaller scales a `hyperbolic' wind geometry may be preferred \citep{2019MNRAS.484.3334S}, in tension with our simulated winds that are bowl-shaped or `parabolic' (in wind geometry, not in particle orbits). This tension may be caused by the lack of infrared radiation pressure in our models -- our winds are initially launched radially rather than vertically.

\subsection{Comparison with dynamical models}

Most previous RHD simulations have mostly been performed in the small-scale regime, where wind generation can be resolved \citep[][Paper I]{2016ApJ...825...67C,2016ApJ...819..115D,2016MNRAS.460..980N,2019ApJ...876..137W}, but below the resolution of most observations. In this work, we have investigated whether the small-scale details are critical for the large-scale (observable) evolution of the wind, and similarly, whether observations can properly constrain these small-scale models. As summarized in Figures~\ref{lum_v}-\ref{lum_r}, we find that large-scale emission is only weakly affected by wind injection parameters, other than varying the extent of emission. Our simulations also showed that an isotropically injected wind can be accelerated into a polar wind by the anisotropy of the accretion disk radiation. Large-scale evolution of outflows is therefore universal, and polar extended dusty gas should be common. This is consistent with the observed polar extension of mid-infrared emission \citep[e.g.][]{2014A&A...563A..82T,2019MNRAS.489.2177A}.

Small-scale effects are however necessary for producing the angle-dependent extinction -- the outflows are only a secondary contribution to opacity. This small-scale extinction can be produced by infrared radiation pressure `puffing up' the disk into a torus \citep{2007ApJ...661...52K,2016ApJ...825...67C,2016ApJ...819..115D,2016MNRAS.460..980N}.

These models neglect supernova feedback, which may play an important role in thickening the disk and driving an outflow \citep{2016ApJ...828L..19W}. Supernova feedback is a complex process, and includes many adjustable parameters, and we will investigate this more fully in a future paper.

The simulations most directly comparable to those of this paper are the larger scale ($32$ pc) simulations performed by \citet{2015ApJ...812...82W,2016ApJ...828L..19W}. A broadly similar fountain structure is formed, with vertical outflow and either weaker winds or inflow at the more equatorial angles. They also find thermally stratified structures, with temperature decreasing from the disk towards the vertical outflow.

One key difference is that they find no outflows are produced when $\gamma_{edd}=0.01$ \citep{2015ApJ...812...82W}. However, as noted in Paper I and above, the radiative acceleration of dense gas is suppressed if the resolution elements are optically thick to AGN radiation. From an examination of their parameters and figures, it appears that this is indeed the case for their disk gas, and that wind production is likely suppressed in their simulations. This further justifies our choice of using an injected wind for our own large-scale simulations.

\subsection{Comparison of wind injection properties with previous RHD models}\label{comparevels}

\begin{table*}
\begin{center}
\begin{tabular}{rcccc}
\hline\hline
Citation & $v_\mathrm{out}$ & $\dot{M}$ & $\dot{p}$ & $\dot{K}$ \\
~ & (km s$^{-1}$) & (M$_\odot$ yr$^{-1}$) & (M$_\odot$ yr$^{-1}$ km s$^{-1}$) & (M$_\odot$ yr$^{-1}$ km$^2$ s$^{-2}$)\\
\hline
\citet{2016ApJ...825...67C} & 1580 & 0.003  & 4.7 & 7500  \\
\citet{2016ApJ...819..115D} (upper limit) & 100 & 0.01 & 1 & 100  \\
\citet{2016MNRAS.460..980N} (min) & 40 & 0.0003 & 0.012 & 0.48 \\
\citet{2016MNRAS.460..980N} (max) & 570 & 0.0007 & 0.4 & 230 \\
\citet{2017ApJ...843...58C} (IR) & 600 & 0.0047 & 2.8 & 1700 \\
\citet{2017ApJ...843...58C} (UV) & 190 & 0.022 & 4.2 & 790 \\
\citet{2019ApJ...876..137W} & 280 & 0.006 & 1.7 & 230 \\
\hline
\end{tabular}
\end{center}
\caption{\label{outflow_vel_table} \textup{Summary of scaled outflow properties in pervious work. Note that these values are measured with different methods and definitions, and can only be coarsely compared.
}}
\end{table*}

We can compare our wind injection parameters to the results of previous high resolution simulations of the inner wind-producing region. To do this, we must rescale their results to the SMBH mass ($M=10^6$ M$_\odot$) and Eddington ratio used here ($\gamma_{edd}=0.01$). We use scalings adapted from equations 34-35 of \citet{2016ApJ...825...67C} by neglecting the opacity-dependent terms, as these terms supply part of the model-dependence that we are testing. The scalings are as follows:

\begin{eqnarray}
v_{out} &\propto& M^{1/4} \gamma_{edd}^{1/4}\\
\dot{M} &\propto& M^{3/4} \gamma_{edd}^{3/4}.
\end{eqnarray}

Using these scaling relations, we extract a wide range of outflow properties from the literature (\citealt{2016ApJ...819..115D,2016ApJ...825...67C,2016MNRAS.460..980N,2017ApJ...843...58C}; paper I). The outflow properties vary with position and time, and so there is no single value for `the' outflow velocity and mass outflow rate of a simulation. Hence the quoted values can not be directly compared in detail, but rather give a sense of some typical values from the simulation.

Mass outflow rates and velocities at infinity (along with scaling relations) are given directly in \citet{2016ApJ...825...67C}. In \citet{2017ApJ...843...58C}, two sets of outflow properties are quoted -- one for the infrared dominated wind, and one for the UV dominated wind, and we show both here. The $\gamma_{edd}=0.01$ model ($\Gamma=0.01$ in their terminology) of \citet{2016ApJ...819..115D} has a late-time mass outflow rate of $\sim10^{-2}$~M$_\odot$yr$^{-1}$, and maximum vertical and radial velocities (at different radii) of $240$~km~s$^{-1}$ and of $220$ kms$^{-1}$. We characterize this simulation with a mass outflow rate of exactly $10^{-2}$~M$_\odot$yr$^{-1}$, and a velocity of  $330$~km~s$^{-1}\sim\sqrt{240^2+220^2}$~km~s$^{-1}$. This outflow velocity represents an upper limit from their simulation. \citet{2016MNRAS.460..980N} quote a mass outflow rate of $0.05-0.1$~M$_\odot$yr$^{-1}$, and velocities of $200-3000$~km~s$^{-1}$, and we use the maximum and minimum of both properties. For our own previous simulation (paper I) we directly re-examine the simulation data for run~a2\_e01 to determine the outflow properties.

The scaled outflow properties are summarized in Table~\ref{outflow_vel_table}. As noted in Section~\ref{section_simulations}, our simulations span the lower range of outflow velocities found in smaller scale RHD simulations, which is the most dynamically interesting regime.

\section{Summary \& Conclusions}\label{section_conclusion}

We performed RHD simulations of and dusty AGN disk (`torus') and wind, to capture evolution on a $\sim1-100$ pc scale ($\gg r_{sub}=0.15$ pc). We varied the parameters of the injected wind, as well as the sub-grid parameters of the AGN. Here we summarize our findings:

\begin{itemize}

\item We find polar extended infrared emission to be almost ubiquitous, as found in observations \citep{2012ApJ...755..149H,2013ApJ...771...87H,2014A&A...563A..82T,2016A&A...591A..47L,2018ApJ...862...17L}. An outflow can evolve into a polar wind after being accelerated by the anisotropic radiation of the AGN accretion disk. Provided the initial outflow speed does not vastly exceed the escape velocity, the equatorial component of the wind tends to `fail', as it receives less radiation pressure from the accretion disk. The vertical component always persists, even with near-zero initial outflow speeds. Vertical winds therefore do not strongly depend on the initial wind launching mechanism, and polar extensions of dusty gas should be expected to be common.

\item We produce a hollow-cone structure, consistent with results from RT models of observations. Our models produce a parabolic cone, while some RT models imply a hyperbolic structure, which may require the inclusion of infrared radiation pressure from the AGN's dusty disk. Overall, our 3D RHD simulations provide the physical justification for the use of disk+wind RT models that show a universally better match to observed SEDs and interferometry than classical torus models \citep{2017ApJ...838L..20H,2017MNRAS.472.3854S,2019MNRAS.484.3334S,2019ApJ...884...10G,2019ApJ...884...11G}.

\item The optical depth of the outflows beyond $r\gtrsim50r_{sub}$ is too thin to explain the observed angle-dependent obscuration \citep{1993ARA&A..31..473A}, which must be produced on a smaller scale -- closer to the sublimation radius -- through infrared radiation pressure.

\item The velocity maps of the simulations are complex, even when the simulation geometry is fairly simple. We reproduce `counter-rotating' features, as found in observations \citep{2019A&A...632A..61G,2019ApJ...884L..28I}.
\end{itemize}

\vspace*{\baselineskip}

In future work we will examine the production of clumps, the role of supernova feedback, and perform post-processing analysis to compare these results to higher mass AGNs such as NGC1068, as well as implement a form of infrared radiation pressure. We will also perform a fuller parameter study to further investigate the effects of varying the mass and Eddington factor.

\section*{Acknowledgements}This research is supported by European Research Council Starting Grant ERC-StG-677117 DUST-IN-THE-WIND.
\bibliographystyle{aasjournal}
\bibliography{rhd_torus_big_1}
\end{document}